\documentclass[12pt]{article}%
\pdfoutput=1
\usepackage{amsmath}
\usepackage{graphicx}
\usepackage[colorinlistoftodos]{todonotes}
\usepackage[colorlinks=true, allcolors=blue]{hyperref}

\usepackage{subcaption}
\usepackage{xr}
\usepackage{changepage}
\usepackage{amsfonts}

\usepackage{amsthm}
\usepackage{amssymb}
\usepackage{natbib}
\usepackage{booktabs}
\usepackage{xr}
\usepackage{multirow}
\usepackage{color}
\usepackage{relsize}
\usepackage{titling}
\usepackage[english]{babel}
\usepackage[utf8x]{inputenc}
\usepackage[T1]{fontenc}
\usepackage{setspace}
\usepackage{thmtools}

\usepackage{lscape} 
\usepackage{placeins}
\providecommand{\U}[1]{\protect\rule{.1in}{.1in}}
\newtheorem{assumption}{Assumption}
{\theoremstyle{definition}
\newtheorem{definition}{Definition}

\newtheorem{observation}{Observation}
\newtheorem{remark}{Remark}
}

\newtheorem{lemma}{Lemma}

\newcommand{\bx}{{x}}
\newcommand{\bX}{{X}	}
\newcommand{\by}{{y}}
\renewcommand{\Re}{\mathbb{R}}
\newcommand{\seps}{\sigma^2}

\newcommand{\argmin}{\operatornamewithlimits{argmin}}
\newcommand{\loss}{L}

\usepackage{scalerel,stackengine}
\stackMath
\renewcommand\widehat[1]{%
\savestack{\tmpbox}{\stretchto{%
  \scaleto{%
    \scalerel*[\widthof{\ensuremath{#1}}]{\kern-.6pt\bigwedge\kern-.6pt}%
    {\rule[-\textheight/2]{1ex}{\textheight}}%WIDTH-LIMITED BIG WEDGE
  }{\textheight}% 
}{0.5ex}}%
\stackon[1pt]{#1}{\tmpbox}%
}

\usepackage{csvsimple}
\usepackage{lscape}
\usepackage{longtable, booktabs,array}
\usepackage[margin=1in]{geometry}
\usepackage[round-mode=places, round-integer-to-decimal, round-precision=2,
            table-format = 1.2,
            table-number-alignment=center,
            round-integer-to-decimal]{siunitx}

\renewcommand{\thmcontinues}[1]{continued}
\begin{document}

\title{\vspace{-1in}\textsc{Competing Models}\thanks{We thank Sylvain Chassang,  Kfir Eliaz, Ben Golub, Yuhta Ishii, Annie Liang, Jonathan Libgober, George Mailath, Stephen Morris, Wolfgang Pesendorfer, David Pearce, Luciano Pomatto, Rani Spiegler, Andrei Shleifer, Stefanie Stancheva, four anonymous referees, and the participants of many seminars and conferences for their useful comments and
suggestions. We thank Stefano Giglio for helpful discussions and feedback. Dibya Mishra, Ken Teoh and Amilcar Velez provided excellent research assistance. Pai gratefully acknowledges the financial support of NSF Grant CCF-1763349.}}
\author{
\small{Jos\'e Luis Montiel Olea\thanks{Department of Economics, Columbia University. Email: \href{mailto:jm4474@columbia.edu}{jm4474@columbia.edu}}, Pietro Ortoleva\thanks{Department of Economics and SPIA, Princeton University. Email: \href{mailto:pietro.ortoleva@princeton.edu}{pietro.ortoleva@princeton.edu}.}, \ Mallesh Pai\thanks{Department of Economics, Rice University. Email: \href{mailto:mallesh.pai@rice.edu}{mallesh.pai@rice.edu}.}, \ Andrea Prat\thanks{Columbia Business School and Department of Economics, Columbia University.\\ \hphantom{\P}Email: \href{mailto:andrea.prat@columbia.edu}{andrea.prat@columbia.edu}}}}
\date{\footnotesize{\footnotesize{\textit{First version}: October 2017} \\\textit{This version}: \today} \\ }

\pretitle{\begin{flushleft}\LARGE} % makes document title flush right
\posttitle{\end{flushleft}}
\preauthor{\begin{flushleft}\large} % makes author flush right
\postauthor{\end{flushleft}}
\predate{\begin{flushleft}\large} % makes date title flush right
\postdate{\end{flushleft}}
\settowidth{\thanksmarkwidth}{*}
\setlength{\thanksmargin}{-\thanksmarkwidth}

\maketitle

\renewenvironment{abstract}
  {\small%\quotation
  {\bfseries\noindent{\abstractname}\par\nobreak\smallskip}}
  %{\endquotation}
  
\begin{abstract}
\noindent 
Different agents need to make a prediction. They observe identical data, but have different \emph{models}: they predict using different explanatory variables. We study which agent believes they have the best predictive ability---as measured by the  smallest subjective posterior mean squared prediction error---and show how it depends on the sample size. With small samples, we present results suggesting it is an agent using a  \emph{low-dimensional} model.  With large samples, it is generally an agent with a  \emph{high-dimensional} model, possibly including irrelevant variables, but never excluding relevant ones. We apply our results to  characterize the winning model in an auction of productive assets, to argue that entrepreneurs and investors with simple models will be over-represented in new sectors, and to understand the  proliferation of ``factors'' that explain the cross-sectional variation of expected stock returns in the asset-pricing literature. 
\end{abstract}
\thispagestyle{empty}

%\vskip 0.2in

%\noindent \textbf{Keywords:} Misspecified Models, Model selection.

%\vskip 0.2in

%\noindent \textbf{JEL:} D44, D83, D03.

\newpage \renewcommand{\thefootnote}{\arabic{footnote}} \pagebreak %
\setcounter{page}{1}

\newpage

\section{Introduction}

The value that individuals assign to a choice often depends on how well they believe they can predict unknown variables. How much an entrepreneur is willing to pay for a company, or whether they choose to enter a new market, depends on their belief in their own ability to predict and respond to future conditions, like market demand, costs, and competition. Households are more likely to invest in financial assets if they believe they can predict future market values. 

In this paper, we study how individuals' assessments of their own predictive ability interacts with the models they use and the available sample size. Our agents are Bayesian and observe the same data, but their predictions are based on different models: some agents believe only a few covariates matter for predictions, while others believe that many more do. We ask: What are the characteristics of the model of the agent who, after observing the data, believes they have the best predictive ability, as measured by the  smallest subjective posterior mean squared prediction error (subjective MSPE)? Colloquially, a candidate who believes they have the best predictive ability may be described as the most ``confident.'’ Similarly, if the subjective MSPE is below the (unknown) objective MSPE of their model, they may be described as ``overconfident.'’ In what follows, we refer to the agent's assessment by subjective MSPE, but in our applications we expand on this confidence/ overconfidence interpretation to deliver novel implications to the behavioral literature on overconfidence.

We show that the answer depends on the model's dimension and the sample size.  With \textit{small samples}, agents with the smallest subjective MSPE use a \emph{low-dimensional} model, using only a few covariates, regardless of the true data generating process (DGP). In contrast, with \emph{large samples}, agents with the smallest subjective MSPE use a \emph{high-dimensional model}, possibly including irrelevant covariates, but never excluding relevant ones. In single-agent decision problems, this results in novel comparative statics: the dimension of agents' models and the dataset's sample size influence the value they assign to each action, holding fixed other standard considerations (e.g., risk aversion, outside options, etc.). In settings where agents compete and relative subjective expected prediction error matters, model dimension and sample size determine the winning model.

\paragraph{Our model.} As a concrete example, consider a second-price auction where a productive asset is sold to the highest bidder. The new owner of the asset will choose an action $a$, and her payoff will be given by $M - (a-y)^2$, where $M$ is a known positive quantity and $y$ is a random variable. Thus, the value of the asset depends on how well the agent can predict $y$. 

There are multiple interesting economic issues in such a setting. Bidders may have different payoff functions, different actions sets, or different information. We abstract from all those issues and focus on the impact of using priors that involve a simpler as compared to a more complex relationship between explanatory variables and the variable of interest $y$.
Specifically, suppose all agents agree that $y$ is a linear function of a number of  covariates $\{x_j\}_{j \in \{1, \dots, k\}}$ plus a noise term, i.e., $y = \sum \beta_j x_j  + \epsilon$. Both the $\beta_j$'s and the variance of $\epsilon$ are unknown, and agents may have different prior distributions on them. In particular, some agents may believe that only a subset of the covariates matters for predicting $y$.

All agents are given the same data: $n$ independent draws of $x$ and $y$, according to an unknown process. Agents are Bayesian but have different priors---as in \cite{harrison1978speculative} or \cite{morris1994trade}. Thus, each agent computes a posterior distribution of $\beta_j$'s and the variance of $\epsilon$, and will use these posterior distributions to solve for their optimal action. Notice that there is no winner's curse in this setup, as winning the auction has no effect on the winner's posterior distribution. Therefore, the auction has the usual equilibrium in dominant strategies where each agent bids her expected value of the asset if she becomes the owner and gets to choose $a$. The winner is the agent with the lowest subjective MSPE. As everything else is equal, this is a competition among models. We ask: What are the characteristics of the model that, after observing the data, has the lowest subjective MSPE? 

Note that there is a trivial reason why certain models may have lower subjective MSPE: their priors may contain less uncertainty about the world. The most extreme case is when an agent is dogmatic and she has a (right or wrong) deterministic model. That agent would believe she has no prediction error and would always bid $M$ in the auction described above. To focus on more interesting effects, we prove all our results under the assumption that, absent data, all agents have the same expected loss. 

\paragraph{Results.}
Our first result, Lemma \ref{lemma:1agentposteriorloss}, characterizes the subjective MSPE of an agent as a function of their prior and observed data.  We prove a subjective/ Bayesian variant of a standard decomposition to show that subjective MSPE can be written as the sum of two components, which we term 1) \emph{model fit}: the agent's posterior expectation of the variance of the regression residual, $\epsilon$; and 2) \emph{model estimation uncertainty}: the agent's degree of uncertainty about the coefficients in her regression model. Crucially, we show that the latter depends on the model's dimension.
This implies that, while our Bayesian agents use their posteriors to compute the best action and do not explicitly care about the dimension of their model, the dimension affects their subjective MSPE. 

This characterization has two immediate implications, depending on the size of the dataset. Our first set of results pertains to the case of small samples. Here we show that ``model estimation uncertainty'' plays a critical role. While agents who use only a few covariates may have a lower model fit, they will also have a lower model estimation uncertainty, since they have fewer parameters to estimate. One complication with small samples is that the actual realized dataset matters, not just the agent's prior. Assuming that priors take a convenient conjugate form typical in Bayesian linear regression, we show how small sample sizes favor small models---even assuming that all agents have the same prior expectation about their prediction error. First, Proposition \ref{prop:1dataunknownP} shows that when the dataset consists of a single datapoint, the lowest subjective MSPE is of a model that contains a single covariate (regardless of the realized data, the true DGP, or the parameters of the prior). Second, Proposition \ref{prop:highprobwinnerUP} shows that for any fixed sample size and for any true data generating process, lower-dimensional models have a lower subjective MSPE with high probability, as long as the prior on the variance of $\epsilon$ is high enough. Intuitively, with small samples, uncertainty about the parameters of the model plays a crucial role. Smaller models have an advantage since uncertainty about the parameters decreases faster as data accumulates. Even if these smaller models are misspecified, if the sample is small their model fit will not be much lower, meaning that they will have the lowest subjective MSPE. Third, we show that smaller models also have a smaller subjective MSPE when agents all believe they know the variance of the error and this variance is common.  Finally, we show that this is also true when agents need to compute their future expected subjective MSPE before data is realized, but knowing that a data set of size $n$, for any $n$, is going to be revealed before the action is chosen.

Next, we consider the case of large sample size. Here, model estimation uncertainty vanishes: agents will have no uncertainty about their fitted parameters, even if they are using the wrong model. Subjective MSPE is therefore based solely on \emph{model fit}. Proposition \ref{thm:large_n_winner-easy} then shows that models that omit a covariate that is relevant for prediction never prevail. At the same time, we also show that high-dimensional models---those that contain additional covariates irrelevant to the true DGP---may continue to win, even asymptotically.  Even though these high-dimensional models will converge to the true DGP, for any finite sample they remain strictly different. We show that the probability of winning for high-dimensional models remains strictly above zero, even asymptotically. In turn, this shows that the role of priors does not vanish asymptotically: it continues to affect a model's probability of winning, even with arbitrarily large samples.

\paragraph{Applications.} In the main body of the paper, we discuss the case of an auction of a productive asset as a leading example. In Section \ref{sec:application}, we present two additional applications. First, we consider a simple model of entry in which returns depend on prediction error: for example, the decision of an entrepreneur to enter a new sector, or the decision of a household to invest in a risky asset. Conditional on entry, the agent must make further choices. For example, as in classic organizational economics models, the entrepreneur must choose a strategy $a$ that fits the (predicted) state of the world $y$ and the loss function is the quadratic difference between $a$ and $y$, as in, e.g., \cite{marschak1972economic}, \cite{roberts1992economics} or \cite{alonso2008does}. Alternatively, the investor must predict price movements to profitably buy/sell the asset.  Agents may have simple or complex models: they make their forecasts using few or many covariates). The results of this paper provide a novel comparative static: in new sectors/asset classes, we should observe an over representation of investors with ``simple'' models, even when reality is ``complex.'' We connect this to the literature on overconfidence of entrepreneurs and investors. 

Second, we use our framework to understand the proliferation of ``factors'' that explain the cross-sectional variation of expected stock returns in the asset-pricing literature. We argue that the increase in the number of test portfolios used to compute the popular Fama-French cross-sectional regressions mechanically favors asset-pricing models with several factors. Our empirical analysis on the evolution of the ``factor zoo'' can be viewed as a particular instance of a simple model of scientific progress. Early models (when there is little data available) may be overly simple relative to the truth, and become more complicated as samples accumulate. 

%\bigskip 

\paragraph{Related Literature.} Our results sit within the large and growing body of work in economic theory on agents with misspecified models; we defer a full discussion to Section \ref{sec:literature}. One key difference is that in most of the literature, misspecified models are evaluated using their objective performance. Our paper, along with a few contemporaneous or subsequent ones \citep{eliaz2018model, levy2019misspecified, he2020evolutionarily}, focuses instead on agents' \textit{subjective} perception of their prediction error, the key metric in our applications.

Our results may also, at a high level, be reminiscent of model-selection methods in Statistics and Machine Learning, with one big difference: our results emerge as the outcome of competition among Bayesian decision makers using different models. By contrast, the model selection literature proposes and studies techniques to explicitly penalize high-dimensional models. The Bayesian statisticians in our paper cannot discard covariates.

%\bigskip

\paragraph{Outline.} The remainder of the paper is organized as follows. Section \ref{sec:model} outlines the formal model, and characterizes the subjective MSPE of a single agent, the foundation of our results. Section \ref{sec:results} illustrates the key trade-offs under competition with a simple numerical simulation.  Section \ref{sec:smalldata} then presents formal results for the case where the size of the dataset, $n$, is small, while Section \ref{sec:bigdata} considers the case where $n$ is large.  Section \ref{sec:application} studies the applications described above. Section \ref{sec:literature} discusses the related literature, and Section \ref{sec:conclude} concludes. 

\section{Model and Single-Agent Problem} \label{sec:model}

Agents want to predict a real-valued variable $y$. There are $k$ real-valued covariates (or explanatory variables) $x \in \Re^k$. 

\paragraph{Data and Data Generating Process.} Before making a prediction, agents observe a common data set, denoted $D_n$, composed of $n$ i.i.d. draws of $y$ and $x$. We  denote the data as $D_n = (Y,X)$, where $Y \in \Re^n$ and $X \in \Re^{n \times k}$. The assumption that all agents observe the \emph{same} data will be relevant for our applications (for example, in an auction setting, this avoids winner's curse).

A true Data Generating Process (DGP), denoted $\mathbb{P}$, determines the joint distribution of the random variables $y$ and $x$. Most of our results assume only that true distribution of covariates has finite moments of all orders (and a positive definite matrix of second moments). 

\paragraph{Statistical Models.} Agents do not know $\mathbb{P}$ but work with a \emph{statistical model}: a family of plausible joint distributions for $y$ and $x$.   In particular, agents posit a linear relation between $y$ and the covariates $x \in \Re^k$, i.e., conditional on $x$:
\begin{equation} \label{equation:DGP}
y = \bx' \beta + \epsilon, \ \ \ \ \text{where} \ \ \ \ \epsilon | \bx \sim \mathcal{N}_1 (0, \seps),\ \  \beta \in \Re^k, %\nonumber
\end{equation} 
that is, agents assume $y|x$ is a \emph{homoskedastic linear regression with Gaussian errors} and parameters $(\beta,\seps) \in \Re^k \times \Re_+$.\footnote{Because covariates in $x$ can be correlated, our framework allows the agents to consider a wide family of non-linear relations. For example, the non-linear process $y=3\frac{x^3_1}{\sqrt{x_5}}+\epsilon$, can be accommodated by defining a new observable equal to $\frac{x^3_1}{\sqrt{x_5}}$. While not all non-linear processes can be expressed this way, especially since we assume finitely many covariates, good approximations can always be achieved. }$^{,}$\footnote{We use the notation $\mathcal{N}_{k}(\mu,\Sigma)$ to denote a multivariate normal distribution of dimension $k$ with mean $\mu$ and covariance matrix $\Sigma$. See p. 171 of \cite{hogg} for a textbook reference on this convention.} 

Agents also assume that the covariates follow a distribution $P$ which belongs to some parametric family.\footnote{A family of distributions $\mathcal{P}$ is said to be parametric, if its elements are indexed by a finite-dimensional vector. One example is $x \sim \mathcal{N}_k(0,\Sigma)$, where $\Sigma$ is an unknown positive definite matrix.} This may or may not be the correct distribution. We assume that, for any element in this family, the matrix $E_{P}[xx']$ is positive definite and that the random vector $x$ has finite moments of all orders.\footnote{Positive definiteness of the matrix rules out the case in which one covariate is a linear combination of some of the others.} 

Together, $P$, $\beta$, and $\seps$ fully define a joint distribution over $y$ and $x$, which we denote by $Q_{\theta}$, with parameter $\theta : = (\beta,\seps,P)$. 

\paragraph{Different Explanatory Variables.} Different agents may consider different explanatory variables in $x$ as relevant for their prediction. We assume that agents consider at least one explanatory variable in their models. The following notation will be useful. If $\{1,2\ldots, k\}$ label the explanatory variables in $x$, we denote by $J \subseteq \{1,2\ldots, k\}$ the subset that an agent considers possibly relevant for prediction. For a given vector $\beta$, the subvector consisting solely of the components in $J \subseteq \{1,\ldots, k\}$ denote by $\beta_J$. Let  $x_J$ be the analogous subvector of $x$, and $X_J$ the corresponding submatrix of $X$.

\paragraph{Misspecification.} An agent who considers the variables in the set $J$ has the statistical model $\{ Q_{\theta_J} \}_{\theta_J \in \Theta_{J}}$. Here $\Theta_J$ is the set of parameters corresponding to variables $J$, i.e., $\theta_J:= (\beta_J, \sigma^2, P).$ This model is said to be \emph{misspecified} if there is no $\theta_J \in \Theta_J$ for which $Q_{\theta_{J}} = \mathbb{P}$ (and it is correctly specified otherwise). In words, a statistical model is misspecified if it does not contain the true DGP  \citep{kleijn2012bernstein}. When the true DGP $\mathbb{P}$ is also a Gaussian linear regression model as in \eqref{equation:DGP} (which we will assume for some of our results), let $J_0$ denote the covariates with non-zero $\beta$s in the true DGP. Then, note that the model associated with any set of variables $J$ for which $J_0 \not\subseteq J$ is necessarily misspecified.\footnote{At first glance, it might seem reasonable to say that a model $y=\beta_1 x_1 + \epsilon$ need not be misspecified when the model is truly $y=\beta_1 x_1 + \beta_2 x_2 + \epsilon'$, as long as the distributions of $\epsilon$ and $\beta_2 x_2 + \epsilon'$ coincide. But this is ruled out by the assumption that error distributions are restricted to be Gaussian, centered at zero, and independent of covariates.} 

\paragraph{Priors.} Agents are Bayesians. An agent who considers variables $J$ as relevant for prediction has a prior $\pi$ over $\Theta_J$. It will be convenient to denote by $J(\pi)$ the set of variables that an agent with prior $\pi$ considers relevant for prediction. Formally, let $\pi_j$ denote the marginal distribution over $\beta_j$ corresponding to prior $\pi$. If $\delta_0$ denotes a Dirac measure at zero, then 
\begin{align*}
    J(\pi):=\{j \in {1, \dots, k}: \pi_j(\beta_j)\neq \delta_0\}.
\end{align*}
In a slight abuse of terminology, we sometimes use $J \subseteq \{1, \dots, k\}$ to refer to a \emph{model}, which should be understood as the set of explanatory variables that are not exactly equal to zero under the prior $\pi$. Strictly speaking, though, a statistical model refers to the collection of distributions over data given parameters as we have defined above; see \cite{mccullagh2002statistical}.

\paragraph{Actions, Utility, and Optimal Prediction.}

Agents make a prediction of $y$ given covariates $x$. Formally, they construct a prediction function $f$ that maps $x$ into $y$, i.e., $f: \mathbb{R}^k \rightarrow \mathbb{R}.$ They minimize a standard quadratic loss function, equal to the square of the difference between the true $y$ and their forecast $f$, i.e., $(y-f)^2.$ Denote by $\loss (f,\theta)$ the agent's loss under prediction function $f$ if the true DGP is $Q_{\theta}$, i.e.,
 \begin{align} \label{eqn:loss2}
     \loss (f,\theta) := \mathbb{E}_{Q_{\theta}}[(y-f(x))^2].
 \end{align}
 If $\pi$ is the agent's prior over $\theta$ and $D_n$ is the observed data, then characterizing the optimal prediction $f^{*}$ is a standard problem.  The agent chooses $f$ to minimize  $\mathbb{E}_{\pi}[\loss (f,\theta)|D_n]$, which can be rewritten as
\begin{equation}
\mathbb{E}_{\pi} [\seps|D_n] + \mathbb{E}_{\pi} \left[  \mathbb{E}_{P}  [(\bx' \beta - f(\bx))^2] \: | \: D_n \right] . \label{eqn:loss}
 \end{equation}
 %\[ \mathbb{E}_{\pi} \left[ \mathbb{E}_{P}  [(\bx' \beta - f(\bx))^2]  |  D_n \right ] \]
The first term does not depend on $f$. The second term involves the average error incurred  in predicting $x^{\prime} \beta$ using $f(x)$.\footnote{The inner expectation averages over values of $x$. The outer one averages over the values of $\beta$ and $P$.} With standard arguments (i.e., exchanging the order of integration and taking first-order conditions), we can see that this is minimized by
\begin{equation} \label{eqn:bayesopt}
f^{*}_{(\pi, D_n)} (\bx) := \bx' \,\, \mathbb{E}_{\pi}[\beta|D_n] = x_{J(\pi)}^{\prime} \mathbb{E}_{\pi}[\beta_{J(\pi)} | D_n] .
\end{equation}
Thus, a Bayesian decision maker with a posterior $\pi|D_n$, model $J(\pi)$, and a square loss function, forecasts $y$ at $x$ as her Bayesian posterior mean of $x'\beta$. This is a standard result.
 
The agent's posterior loss, conditional on her using the optimal prediction function characterized above, is denoted $L^*(\pi, D_n)$. We refer to it as the \emph{subjective posterior mean-squared prediction error} (subjective MSPE). 

%In what follows, we may omit the qualifier of posterior, or even refer to it as simply prediction error for brevity. 
 
\subsection{Decomposing Subjective MSPE} We now characterize an agent's subjective MSPE, our key dimension of interest. The key forces at play will already be evident from the following lemma. 

\begin{restatable}{lemma}{oneagentposteriorloss}\label{lemma:1agentposteriorloss}
Suppose that $\beta_{J(\pi)}$ is independent of $P$ under the posterior distribution. The agent's subjective MSPE can be decomposed as:
\begin{align}
\label{equation:PosteriorLoss}
&L^*(\pi, D_n) =  \underbrace{\mathbb{E}_{\pi} \left[\seps | D_n\right]}_{\textrm{Model Fit}} + \underbrace{\textup{tr}\left(\mathbb{V}_{\pi}[\beta_{J(\pi)}|D_n] \; \mathbb{E}_{\pi} \left[ \mathbb{E}_{P}[\bx_{J(\pi)}\bx_{J(\pi)}'] \: | \: D_n \right]  \right)}_{\textrm{Model Estimation Uncertainty}},
\end{align}
where $\mathbb{V}(\cdot)$ is the variance-covariance operator, and $\textup{tr}$ is the trace operator. 
\end{restatable}

This lemma is reminiscent of standard decompositions of \emph{mean-squared prediction error} in frequentist linear regression models \cite[e.g., ][Theorem 4.8]{hansen2020econometrics}, except that in this case it characterizes the \textit{subjective} MSPE of the agent using their own prior. The lemma shows that the agent's subjective MSPE, $L^*(\pi,D_n)$, is the sum of two components. The first,  the posterior expectation of $\sigma^2_{\epsilon}$, is the agent's estimate of the irreducible noise in the system. We interpret this term as a measure of  \textit{model fit}, i.e., how well the model explains the data (as all unexplained variation must be ascribed to noise). 

The second term, $\textup{tr}\left(\mathbb{V}_{\pi}[\beta_J|D_n]\; \mathbb{E}_{\pi} \left[ \mathbb{E}_{P}[x_Jx_J'] \: | \: D_n \right] \right)$, is  the trace of the variance-covariance matrix of the coefficients of the model (adjusted by the posterior mean of $\mathbb{E}_{P}[x_Jx_J']$). We interpret this term as a measure of how uncertain the agent is in her estimation of the parameters of the model according to her own prior, capturing \textit{model estimation uncertainty}. To illustrate why, consider the simpler case in which the posterior mean of $\mathbb{E}_{P}[x_J x_J']$ is the identity matrix. Then, the second term reduces to $\textup{tr}\left(\mathbb{V}_{\pi}[\beta_J|D_n]\right)$, i.e., $\sum_{j\in J} \mathbb{V}_{\pi}[\beta_j|D_n]$; this is simply the sum of the posterior variances of the parameters $\beta_j$, indeed a measure of model estimation uncertainty. In the next section we will show that this decomposition has immediate implications for which model leads to the lowest subjective MSPE. 

\begin{remark}\label{ref:rem_independence}
The independence of $\beta_J$ and $P$ under the posterior distribution will hold under general assumptions. For example, it holds when agents know the distribution of covariates, or if we consider statistical models in which $P$ does not enter the parametric model of $y|x$ and $(\beta,\sigma^2)$ does not affect the distribution of $x$.
\end{remark} 

\section{Competing Models: A First Look} \label{sec:results}

Suppose agents participate in a mechanism that selects the agent with the \textit{lowest} subjective MSPE. A leading example, discussed in the introduction, is that of a second-price auction of a productive asset. In this scenario, the winner of the productive asset will choose an action $a$ and her payoff will be given by $M - (a-y)^2$, where $M$ is a known positive quantity and $y$ is the variable agents aim to predict. Thus, the value of the asset depends on how well the agent can predict $y$. Assuming that different priors are the only dimension of heterogeneity among participants, and noting that there is no winner's curse, the auction will select the agent with the lowest subjective MSPE. Applying Lemma \ref{lemma:1agentposteriorloss}, we immediately derive that this is the agent with \emph{the best trade-off between model fit and model estimation uncertainty.} 
% \[ \underset{\pi}{\mathrm{min}} \ \ \underbrace{\mathbb{E}_{\pi} \left[\seps | D_n\right]}_{\textrm{Model Fit}} + \underbrace{\textup{tr}\left(\mathbb{V}_{\pi}[\beta_{J}|D_n] \; \mathbb{E}_{\pi} \left[ \mathbb{E}_{P}[x_Jx_J']  | D_n \right] \right)}_{\textrm{Model Estimation Uncertainty}},\] 
% This is the prior that 
% \emph{achieves the best trade-off between model fit and model estimation uncertainty.} 

Before we dive into formal results, we present a simple simulation to illustrate the key forces at play and our main findings. Suppose that there are six covariates, $\{x_1, \ldots , x_6\}$, of which only the first five are relevant for prediction in the true DGP, i.e., $y = \sum_{j=1}^5 \beta_j x_j + \epsilon$, and $\epsilon | x \sim \mathcal{N}(0,\seps)$. For simplicity, assume that each nonzero regression coefficient, $\beta_j$, is equal to $1$, as is $\seps$. Also assume that $x \sim N(0,\mathbb{I}_6)$ under the true DGP. 

Considering all subsets of covariates, there are $63$ agents with linear regression models, one for each nonempty subset of  $\{x_1, \ldots , x_6\}$. By construction, 61 are misspecified, one has the exactly correct model, and one has a model of higher dimension compared to the true DGP.  For the simulation, we assume that agents' priors belong to a well-behaved family, common in Bayesian linear regression, and parametrize them so that all agents have the same subjective MSPE before seeing any data.\footnote{Specifically, the agents are assumed to have the following parametric model for their covariates: $x_J \sim \mathcal{N}_{|J|}(0, \Sigma_J)$ where $\Sigma_J$ is an unknown, positive definite matrix. Further, we assume that the priors over the $\beta$,$\seps$ and $\Sigma_J$ belong to the Normal-Inverse-Gamma-Inverse-Wishart family of Definition \ref{def:nigprior} below. In this simulation we set $a_0=2$, $b_0=10$, and $\gamma_0 = .001$.}

\begin{figure}[t!]
    \centering
    \includegraphics[height=4in]{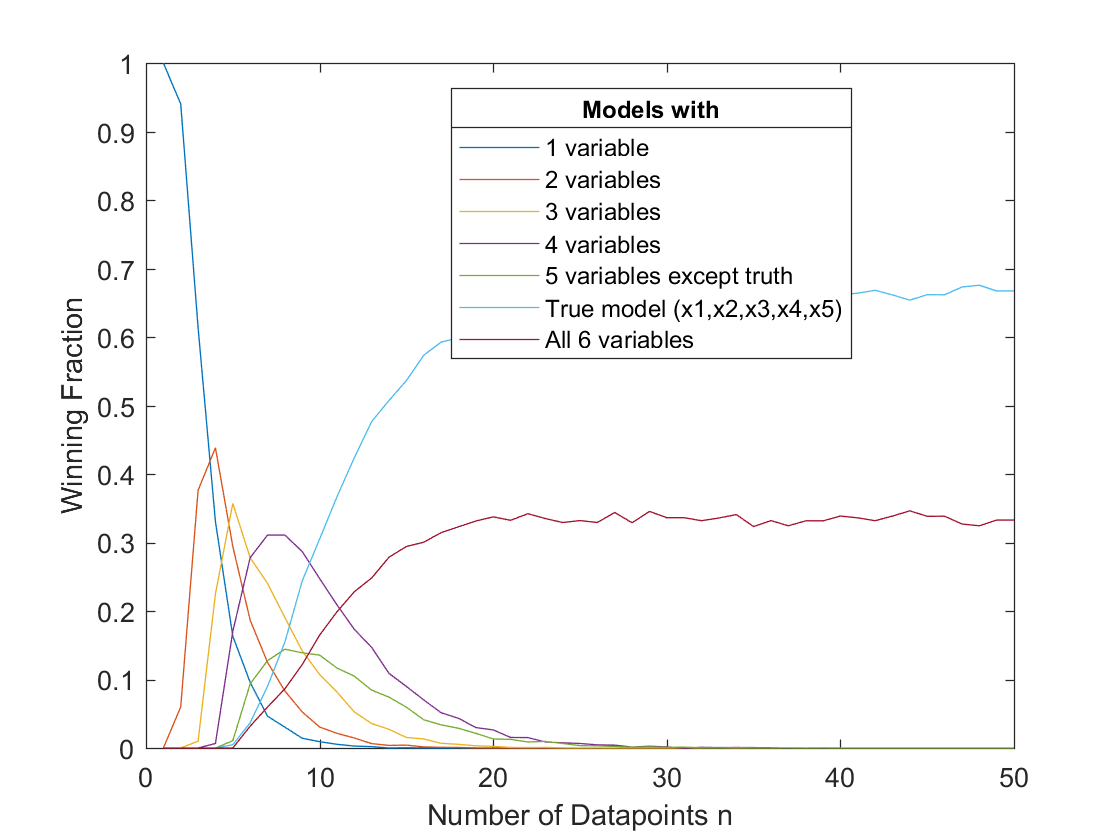}
    \caption{Winning probabilities on $5,000$ simulated datasets of size $n=1, \ldots, 50$.}
    \label{fig:winner}
\end{figure}

Figure \ref{fig:winner} plots the frequency of the size of the model of the agent with the lowest subjective MSPE for datasets of size $n \in \{1, \ldots, 50\}$. Two patterns emerge.

First, when $n$ is small, low-dimensional models tend to win despite being  misspecified. In fact, when $n=1$, we can see in Figure \ref{fig:winner} that the winner is a model with a \textit{single} covariate. 

Second, as $n$ grows large, misspecified models never win. %In Proposition \ref{thm:large_n_winner-easy} we show that this result holds for a very general class of priors. One interesting feature in Figure \ref{fig:winner} is that 
At the same time, the high-dimensional model that includes the redundant variable $x_6$ continues to win with relative frequency that appears to converge to a steady state close to 0.3---strictly above 0. %Proposition \ref{thm:large_n_winner-easy} also shows that this result holds more generally. In fact, in Section \ref{subsection:chi-squared-formula} in the Supplementary Appendix, we present a formula for the probability that a high-dimensional model defeats the true model. Such a formula depends on the prior on $(\beta_J,\seps)$, the parameters of the true DGP, and the models' dimension. Applying our formula to the simulation set-up we find that the probability that a model with 6 covariates defeats the true DGP is approximately the probability that a $\chi^2_1$ random variable exceeds $1-\gamma_0(\beta_0'\beta_0)$. This probability is $0.3161$ (which is close to what we see in Figure \ref{fig:winner}). 

We will now show that both patterns hold more generally.

% Finally, Figure \ref{fig:winner2} reports the winning fraction of models of size $1$, as we increase the shared hyper-parameter $b_0$ (all other simulation parameters stay the same as above). Growing $b_0$ corresponds to a larger prior mean for all agents. The figure shows that, as $b_0$ grows large, the winning probability of low-dimensional models becomes larger, and we formalize this observation in Proposition \ref{prop:highprobwinnerUP}. As we explain later, the rationale for the result comes from the fact that a large $b_0$ implies that the model fit for any two models is essentially the same. Therefore the comparison becomes purely one of model estimation uncertainty, which favors low-dimensional models.  

% \begin{figure}[h!]
%     \centering
%     \includegraphics[height=4in]{figures/varyingb0.png}
%     \caption{Winning rates for models with one covariate as the shared hyper-parameter $b_0$ increases. All other simulation parameters are the same as Figure \ref{fig:winner}.}
%     \label{fig:winner2}
% \end{figure}

\section{The Winner with Small $n$}
\label{sec:smalldata}

We begin with the case in which the number of observations $n$ is small. For tractability, we focus on a special class of priors, widely used in Bayesian linear regression.

\begin{definition}\label{def:nigprior}
A prior has the \textit{Normal-Inverse-Gamma-Inverse-Wishart form} with \emph{hyper-parameters} $(a_0, b_0,\gamma_0)$, with $(a_0, b_0,\gamma_0) \gg 0$ and $a_0 > 1$, if
\begin{align*}
 \beta_{J} | \seps  \sim  \mathcal{N}_{|J|}\Bigg(\textbf{0}, \frac{\seps}{\gamma_0 |J|} \mathbb{I}_{|J|} \Bigg),  \ \ \ \ \ \   \seps  \sim  \textrm{Inv-Gamma}(a_0, b_0), \\  x_{J}   \sim  \mathcal{N}_{|J|}(0,\Sigma_J),  \ \ \ \ \ \   \Sigma_{J}  \sim  \textrm{Inv-Wishart}(\gamma_0 |J| \mathbb{I}_{|J|}, 2|J|+1)
  %\label{prior-normal-invgamma2}
\end{align*}
where $\textrm{Inv-Gamma}(a_0,b_0)$ is the Inverse-Gamma distribution with parameters $a_0$ and $b_0$, and Inv-Wishart($\gamma |J| \mathbb{I}_{|J|}$,$2|J|+1$) is the Inverse-Wishart distribution with $J \times J$ scale matrix $\gamma |J| \mathbb{I}_{|J|}$ and degrees of freedom $2|J|+1$.\footnote{The Inverse-Gamma is a two-parameter family of distributions on the positive real line. For parameters $a_0>1, b_0 \geq 0$ it has mean $b_0/a_0-1$. The Inverse-Wishart is a two-parameter family of distributions on real-valued positive definite matrices. The first parameter (scale matrix) is a symmetric positive definite matrix. The second is a non-negative scalar at least as large as the dimension of the scale matrix.}  The prior on $\Sigma_J$ is assumed independent of $(\beta, \seps)$.
\end{definition} 
The Normal-Inverse-Gamma-Inverse-Wishart priors are conjugate priors for the Gaussian linear regression model and the posterior can be expressed in a closed form as a function of the data---see Appendix \ref{app:proofoflemma2} for details. All results in this section have analogs in the setting where the distribution $P$ on covariates is assumed to be known (not necessarily Gaussian) and $E_{P}[xx']= \mathbb{I}_k$.

One convenient feature of this family of priors, readily checked, is that they imply that all agents have the same subjective MPSE when no data is released, i.e., when $n=0$. Differences in subjective MSPE arise therefore only from the fact that the subjective MSPE evolves differently for models of different dimensions.

Using this family of priors allows us to further simplify the expression for subjective MSPE in Lemma \ref{lemma:1agentposteriorloss}.

\begin{restatable}{lemma}{unknownP}\label{prop:unknownP}
Consider a prior $\pi$ as in Definition \ref{def:nigprior}. Then
\begin{align}\label{eqn:unknownPloss}
    L^*(\pi, D_n) =  \underbrace{\mathbb{E}_{\pi}[\seps|D_n]}_{\textrm{Model Fit}} + \underbrace{\mathbb{E}_{\pi}[\seps|D_n] \left( \frac{|J(\pi)|} {  n + |J(\pi)|} \right). }_{\textrm{Model Estimation Uncertainty}}
\end{align}
%where $E_{\pi}[\seps|D_n]$ is as given in \eqref{equation:PosteriorMeanSigma}.
\end{restatable}

Equation \eqref{eqn:unknownPloss} shows that the dimension of an agent's model enters explicitly into the subjective MSPE via the term $|J(\pi)| \: / \: ( n + |J(\pi)|).$  As this is increasing in $|J(\pi)|$, higher-dimensional models have a disadvantage: as they have more parameters to estimate, their model estimation uncertainty will decrease more slowly. For higher-dimensional models to have lower subjective MSPE, therefore,  they must compensate for this by a sufficiently better model fit. 
Specifically, the ratio between the model fits must be  such that 
\begin{align}\label{eqn:compareUP}
    L^*(\pi, D_n) \geq L^*(\pi',D_n) \iff \frac{\mathbb{E}_{\pi}[\seps|D_n]}{\mathbb{E}_{\pi'}[\seps|D_n]} \geq \frac{\left(1 + \frac{|J(\pi')|}{n + |J(\pi')|}\right)}{\left(1 + \frac{|J(\pi)|}{n + |J(\pi)|}\right)}.
\end{align}
%The right-hand side of the inequality is strictly less than $1$ whenever $|J(\pi')|<|J(\pi)|$. Providing an explicit characterization of model fit is more complicated, as it typically depends on the realized data, and the parameters of the prior. 

In general, however, characterizing the model fit is analytically difficult, as it depends also on the realized data $D_n$, which in small samples can vary substantially. The following results describe how lower-dimensional models have lower subjective MSPE in some cases in which model fit is simple to analyze. First, when the sample size $n=1$; we this to be the case for a model using a single covariate. Second, it holds true when prior variance of $\epsilon$ of the models is high enough relative to $n$. Third, it applies when either all agents believe they know the variance of the error term, $\seps$, and have the same belief, or when subjective MSPE is computed before actual data is released but the agents know that it will be released before they must choose their action---a case that has direct applications (as we will discuss later).

\subsection{One-Dimensional Models Win when $n=1$}

The first result shows that when $n=1$, low-dimensional models have the lowest subjective MPSE, regardless of the realized data, the true DGP, or the parameters of the prior. 

\begin{restatable}{proposition}{onedataunknownP} 
\label{prop:1dataunknownP} 
Suppose all priors are as in Definition \ref{def:nigprior} with shared hyper-parameters. Suppose also that for every single covariate model, i.e., every $J$ such that $|J|=1$, there is an agent with that model, and that all agents use at least one covariate. If $n=1$, then the agent with lowest subject MSPE is an agent with a single covariate model.
\end{restatable}

We prove this result by showing that, when $n=1$, model fit is minimized by some model that considers only a single variable. This means that there is no model with more than one covariate that can improve the model fit of the best one-dimensional model.

\subsection{Low-Dimensional Models Win when Prior Variance Is High}\label{sec:smallnchangeprior}

The sharp characterization obtained for $n=1$ does not hold for other sample sizes. Indeed, our simulations show that models with more than one covariate may have the lowest subjective MSPE with positive probability for $n>1$. As we have discussed, the identity of this model depends on the trade-off between model fit and model estimation uncertainty.

One way to capture the advantage of small-dimensional models when $n$ is small is the following. One characteristic of few data points is that the prior continues to play a relevant role. This can be captured by making sure that the prior mean is large enough relative to the amount of data. In this case, high-dimensional models cannot improve sufficiently on the model fit term relative to lower-dimensional models (as the model fit is poor for any model). Therefore, in this setting also, the advantage that low-dimensional models have in terms of model uncertainty is the dominant factor. %Formally, we have the following result: 

\begin{restatable}{proposition}{highprobwinnerUP}
\label{prop:highprobwinnerUP} 
Let $\Pi$ be a finite set of agents' priors that satisfy Definitions \ref{def:nigprior} with shared hyper-parameters $(a_0,b_0, \gamma)$. Let $|\underline{J}|$ be the size of the smallest model in this set.  Fix the size of the dataset $n$. For any $p \in (0,1)$, there exists $b_0$ large enough so
\begin{align*}
    \mathbb{P}\left(D_n : \exists \pi^* \in \argmin_{\pi \in \Pi} L^*(\pi,D_n) \text{  s.t. } |J(\pi^*)| = |\underline{J}|  \right) >p,
\end{align*}
i.e., with probability at least $p$ over datasets $D_n$, the agent with the lowest subjective MSPE has the smallest size model among all the agents.
\end{restatable}

\begin{figure}[!ht]
     \centering
     \includegraphics[height=4in]{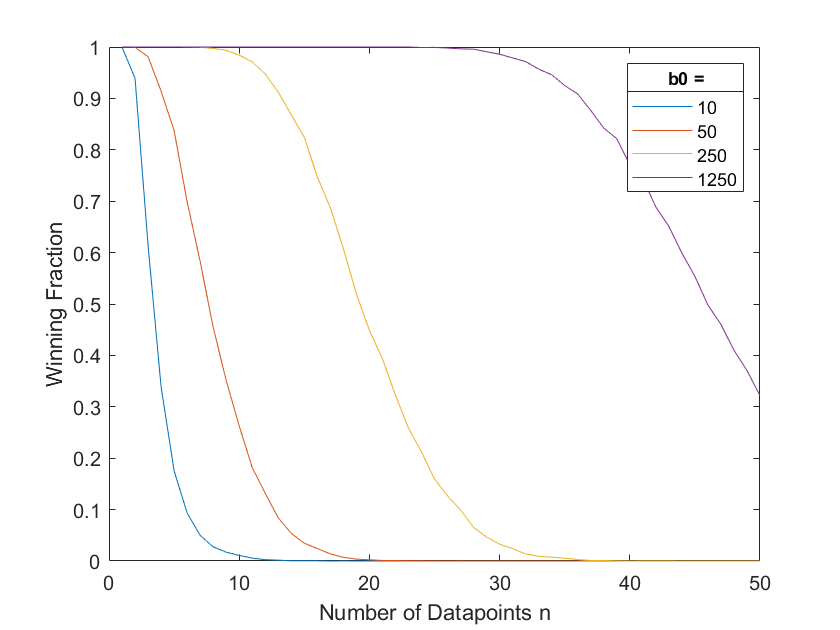}
     \caption{Winning rates for models with one covariate as the shared hyper-parameter $b_0$ increases. All other simulation parameters are the same as Figure \ref{fig:winner}.}
     \label{fig:winner2}
     \end{figure}

As an illustration of this result, let us return to the simulations in Section \ref{sec:results}. Figure \ref{fig:winner2} reports the winning fraction of models of size $1$, as we increase the shared hyper-parameter $b_0$ (all other simulation parameters stay the same as above). Growing $b_0$ corresponds to a larger prior mean for all agents. 

\subsection{Low-Dimensional Models Win when Error Variance Is Known or Data Are Not Released but Expected}\label{sec:smallnchangeprior}

We conclude this analysis considering two other cases in which the model fit is easy to solve analytically, which again show an advantage of lower-dimensional models. We present them as two observations since they follow directly from Lemma \ref{prop:unknownP} above.

%We have the following two straightforward observations about the role played by the uncertainty in $\seps$.

\begin{observation}
Suppose all agents treat $\seps$ as a known common value, but the priors on $\beta|\seps$ and $\Sigma$ are as in Definition \ref{def:nigprior}. Then, for any $D_n$,
$$
L^*(\pi', D_n) < L^*(\pi,D_n) \ \Longleftrightarrow \ J(\pi')|< |J(\pi)|.
$$
That is, if $\seps$ is believed to be known (possibly incorrectly), subjective MSPE is ranked by model dimension.
\end{observation}

This result shows that when all agents believe they know the error variance $\seps$, then model dimension induces a precise ranking between models: smaller models \textit{always} have smaller subjective MSPE. This result follows directly from Lemma \ref{prop:unknownP}.

We now turn to the case in which agents have not received any data, but know that they will receive $n$ data points before making their choice of action. They therefore have to compute their expected subjective MSPE, which we denote by $\mathbb{E}_{\pi'}[L^*(\pi', D_n)]$.

\begin{observation}
For any $\pi$ and $\pi'$, 
$$
\mathbb{E}_{\pi'}[L^*(\pi', D_n)] < \mathbb{E}_{\pi}[L^*(\pi,D_n)] \ \Longleftrightarrow \ J(\pi')|< |J(\pi)|.
$$
\end{observation}

This result shows that if agents have not received any data but know that they will receive it later, again we find small-dimensional models have lower expected MSPE. This observation also  follows straightforwardly from Lemma \ref{prop:unknownP}: by the martingale property of beliefs, the expected model fit is constant among models, meaning that the winner must be a low-dimensional model.

\section{The Winner with Large $n$} 
\label{sec:bigdata}

We now characterize the winner for large $n$. Our results will be derived for a much more general class of priors than the previous section, but it is helpful to start by recalling Lemma \ref{prop:unknownP}, which assumes Normal-Inverse-Gamma-Inverse-Wishart priors. In this case, the subjective MSPE is
\[ \underbrace{\mathbb{E}_{\pi}[\seps|D_n]}_{\textrm{Model Fit}}  + \underbrace{\mathbb{E}_{\pi}[\seps|D_n] \frac{|J(\pi)|}{n + |J(\pi)|)}}_{\textrm{Model Estimation Uncertainty}}.\]
From this formula, it is immediate to see that model estimation uncertainty vanishes as $n$ grows large, making the model fit the crucial aspect. This means that, because misspecified models have worse model fit than correctly specified ones, they must therefore also have worse subjective MPSE when $n$ is large enough.

The comparison is, however, less straightforward between a model that uses the exact same variables as the true DGP and another that also includes additional irrelevant covariates. For both, model fit converges to the true residual variance ($\sigma^2_0$), since both are correctly specified, and model estimation uncertainty converges to zero. Which one has lower subjective MSPE depends on how quickly these converge, which in turn depends on the realized data and on the prior. Our simulations suggest that the long-run behaviors may be such that the larger model may continue to win, with a probability bounded away from zero even at the limit. We will now show how this holds in general.

For this analysis, we do not need to assume that priors have a specific form as we did in the previous section. We simplify our analysis in the body of the paper by making two assumptions: that the true DGP is of the linear Gaussian form, which allows some models to be identical to the true DGP (Assumption \ref{ass: high-level-easy}); and that priors over the $\beta_i$s have full support with a smooth density (on the subset of relevant covariates $J(\pi)$), while priors on $\sigma^2_{\epsilon}$ are not degenerate (Assumption \ref{ass:prior-normalize-easy}).

\begin{assumption} \label{ass: high-level-easy}
There exist parameters $\theta_0 := (\beta_0, \sigma_0^2,P_0)$ such that $Q_{\theta_0} = \mathbb{P}$.
\end{assumption}

\begin{assumption} \label{ass:prior-normalize-easy}
Priors are characterized by a smooth and strictly positive probability density function $\pi(\cdot)$ over $({\beta_{J(\pi)}}^{\prime},\seps)^{\prime} \in \mathbb{R}^{|J(\pi)| }\times \mathbb{R}_+$. The prior over $({\beta_{J(\pi)}}^{\prime},\seps)^{\prime}$ is independent of the prior over $P$.\footnote{By definition, the prior of an agent for any $\beta_{\kappa}$, $\kappa \not\in J(\pi),$ is degenerate at 0.} In addition, for each agent, there exists $n$ large enough for which   $\mathbb{E}_{\pi}[\seps | D_n] < \infty$ almost surely. \end{assumption}

Recall that $J_0$ denotes the set of covariates that are relevant in the true DGP.
\begin{restatable}{proposition}{largenwinnereasy}\label{thm:large_n_winner-easy}
Suppose the true DGP $\mathbb{P}$ satisfies Assumption \ref{ass: high-level-easy} with parameters $(\beta_0, \sigma_0^2)$. Let $\Pi$ be a finite collection of priors that satisfy  Assumption \ref{ass:prior-normalize-easy} and that contains $\pi^*$ with $J_0 \subseteq J(\pi^*)$. If 
\begin{align}\label{eqn:condition} \textrm{tr} \left( n \mathbb{V}_{\pi}(\beta_{J(\pi)}  |  D_n  ) \mathbb{E}_{\pi} \left[ \mathbb{E}_{P}[ x_{J(\pi)} x_{J(\pi)}' ] \: | \:  D_n \right] \right) = O_{\mathbb{P}}(1),
\end{align}
for every prior $\pi \in \Pi$, then
$$\lim_{n \rightarrow \infty} \mathbb{P} \left(  \exists \pi \in \argmin_{\pi \in \Pi} L^*(\pi, D_n)  \textrm{ s.t }  J_0 \nsubseteq J(\pi)      \right) = 0.$$ 
Moreover, for any $\pi$ for which $J_0 \subset J(\pi)$,
$$ \lim_{n \rightarrow \infty} \mathbb{P}\big( L^*(\pi,D_n) < L^*(\pi_0,D_n) \big) \in (0,1],$$
where $\pi_0$ is any prior for which $J(\pi_0) = J_0$.
\end{restatable}

A crucial assumption in Proposition \ref{thm:large_n_winner-easy} is \eqref{eqn:condition}. 
This assumption will be verified whenever the posterior variance of $\beta$ decreases to zero at rate $n$. Lemma \ref{prop:unknownP} already tells us that this condition is satisfied in the special case of Normal-Inverse-Gamma-Inverse-Wishart priors. In fact, this condition holds very generally due to the Bernstein-von Mises theorem, which states that posterior distributions based on parametric models (misspecified or correctly specified) will typically behave like Gaussian distributions, with a variance that decreases at rate $n$.\footnote{See the Bernstein-von Mises theorem for misspecified parametric models of \cite{kleijn2012bernstein}. This result can be thought of as richer versions of the classical results concerning posterior distributions of misspecified models in \cite{berk1970consistency}.} 

Proposition \ref{thm:large_n_winner-easy} has two takeaways. The first part tells us that a misspecified model, because it excludes relevant variables, never wins as the sample size grows large. Any model that is not misspecified will have lower subjective MPSE with probability approaching $1$ as $n$ grows large. The second part shows that \textit{any} model larger than the true one defeats the latter with a  probability that is \textit{strictly positive}, even asymptotically.

We have already discussed the intuition for the first result. The assumptions of the theorem, i.e. Assumptions \ref{ass: high-level-easy} and \eqref{ass:prior-normalize-easy} combined with \eqref{eqn:condition},  guarantee that model estimation uncertainty converges to zero for all agents. 

For the second result, from a technical perspective, our result is based on an  asymptotic expansion for the posterior mean of the variance parameter in the linear
regression model based on the general results in  \cite{KTK:1990}. This is not a fairly technical result, so, for some intuition, let us return to the case of  Normal-Inverse-Gamma-Inverse-Wishart priors. Here, when $n$ is large, it is possible to approximate its distribution using asymptotic theory.

To this end, let $\widehat{\beta}_J$ denote the OLS estimator based on the variables listed in $J$, and let $\widehat{\sigma}^2_J$ be the corresponding residual variance estimator: 
 \[ \widehat{\sigma}^2_J \equiv  (Y-X_J'\widehat{\beta}_J)'(Y-X_J'\widehat{\beta}_J)/n.\] 
A key observation in our analysis---that holds in the Normal-Inverse-Gamma model, but also for more general priors---is that as the sample size grows large
 \begin{equation}
         n \left( E_{\pi_0}[ \sigma^2 | D_n  ] - E_{\pi}[ \sigma^2 | D_n  ] \right) = n \left(\widehat{\sigma}^2_{J_0}-\widehat{\sigma}^2_J \right) + O_{\mathbb{P}}(1),
 \end{equation}
 where $O_{\mathbb{P}}(1)$ refers to a term that is bounded with high probability under $\mathbb{P}$. In the case of Normal-Inverse-Gamma-Inverse-Wishart priors, algebraic manipulations can be used to verify the approximation with a leading term equal to   
 \[ -\gamma \beta_0'\beta_0 (|J|-|J_0|).\] 
 Deriving an analogous result for other priors requires additional effort, given the lack of closed-form solutions for the posterior distributions.\footnote{We refer the reader to Lemma \ref{lemma:expansion}, which uses the Kaas-Tierney-Kadane expansions of posterior moments in \cite{KTK:1990} to verify the approximation.} 

Assumption \ref{ass: high-level-easy}, along with standard results from regression analysis---e.g., Equation 5.28 in \cite{greene2018econometric} and Theorem 5.1 therein---implies $n(\widehat{\sigma}^2_{J_0}-\widehat{\sigma}^2_J)/\sigma^2_0$ converges in distribution to a chi-squared random variable with $|J|-|J_0|$ degrees of freedom. 
This means that the probability that the larger model wins can be approximated by the probability of the event:
 \[  \chi^2_{|J|-|J_0|} -\gamma (\beta_0'\beta_0/\sigma^2_0) (|J|-|J_0|) >  (|J|-|J_0|).  \]

In the context of our simulations---where the true DGP only included the first five covariates, with coefficients $\beta=(1,1,1,1,1)'$, and $\seps=1$---the probability that a model with six variables defeats the true DGP is roughly 
 \[ P(\chi^2_{1}  > 1 + 5 \gamma).  \]
 When $\gamma=.001$, this probability is $0.3161$, which is close to what we see in Figure \ref{fig:winner}. We provide a more general formula in Appendix \ref{subsection:chi-squared-formula}.

It is important to remark that these results hold even if the true DGP is different from \eqref{equation:DGP}. For example, the distribution of errors in the true DGP may be heteroskedastic or non-normal with thin-enough tails, the distribution on covariates $P$ may be misspecified as long as it has finite second moments, and the true DGP may not be linear. Although Proposition \ref{thm:large_n_winner-easy} is presented under special conditions (Assumptions \ref{ass: high-level-easy} and \ref{ass:prior-normalize-easy}), it is possible to prove that our results continue to hold under much weaker ones.\footnote{For example, for the expansion of Lemma \ref{lemma:expansion} to hold, priors do not need to be smooth. It is sufficient that they are differentiable up to the fourth-order.} We hope that the simpler framework helps the reader understand the main forces at play in the competition among models.

\paragraph{Connection with the Akaike Information Criterion.}
A different way to understand our results is to relate the model selection they induce to the Akaike Information Criterion (AIC), a well-studied model selection criterion in econometrics and statistics. 
In what follows, we illustrate that the loss function of an agent with Normal-Inverse-Gamma-Inverse-Wishart prior is ``close'' to the AIC for the linear regression model.

\begin{definition}[Akaike Information Criterion]
Given a dataset $D_n = (Y, X)$ with $n$ data points and $k$ possible covariates, the AIC for linear regression evaluates a model $J$ as
\begin{align*}
&L_{\text{Akaike}} (J,n, D_n) = \ln \widehat{\sigma}^2_J + \frac{2|J|}{n},\\
\text{where }
&\widehat{\sigma}^2_J = \frac1n \min_{\beta \in \Re^{|J|}} (\by - \bX_J \beta)' (\by- X_J \beta). 
\end{align*}
\end{definition}

The expression $\widehat{\sigma}^2_J$ is the OLS estimator of the residual variance based on a model with covariates $X_J$ in the dataset $D_n$. As is well understood, the model with lower estimated variance may not be the model with the best out-of-sample performance. This is because selecting based on average residuals favors models that have more covariates, which may overfit the data. The AIC compensates for this by adding a penalty term equal to $\frac{2|J|}{n},$ i.e., twice the ratio of the number of covariates in the model and the number of data points. Algebra shows that, if agents have an uninformative Normal-Inverse-Gamma-Inverse-Wishart prior, then the posterior loss is approximately equal to 
\[ \ln \Bigg( \widehat{\sigma}^2_J \Bigg) + \ln \left(1 + \frac{|J|}{|J|+n}  \right). \]

Thus, if the sample size is large and the agents' distribution of covariates is well-specified, the posterior loss of an agent with prior $\pi$ is approximately equal to the AIC (with a penalty of $\ln (1 + |J|/(|J|+ n)) \approx |J|/n$ instead of $2|J|/n$). 

The prevalence of larger models in the model competition can be then associated to the ``conservativeness'' of the AIC for model selection. Proposition \ref{thm:large_n_winner-easy}, however, makes clear that the relation is only qualitative: larger models will prevail in large samples, but the probability of a larger model being selected will continue to be affected by the prior.  

Finally, it is worth reiterating that the foundations of the AIC are normative: the criterion was proposed as a way to select models to avoid overfitting. Conversely, our analyis provides a \emph{positive} foundation for a solution similar to the AIC: we study the outcomes when Bayesian agents compete in a way that selects the agent with the lowest subjective MPSE. 

\section{Applications}\label{sec:application}

In previous sections, we considered the auction of a productive asset as a leading example. We now discuss two additional applications. The first is a simple model of entry when returns depend on prediction error, which we connect to the literature on overconfidence. Second, we use our framework to understand the proliferation of ``factors'' in the asset pricing literature.

\subsection{Selection of Simple Models in Entry and Investment}
We begin with an application to a single-agent decision problem. An agent is faced with a risky entry choice---she has to choose between a risky option and a safe one that gives her a utility normalized to $0$. The utility of the risky option depends (in part) on the agent's ability to predict an unknown variable and take an action. This is the case of an entrepreneur who has the option to invest in a new venture, where expected returns depend in part on the ability to predict and adapt to future market demand, political situations, or trade agreements. Alternatively, this could be an individual investor considering trading an asset: returns depend  on the ability of the investor to predict future price movements and trade accordingly.

Formally, suppose that the expected utility of the risky option is
$$
\mathbb{E}(r)=v - L^{*}(\pi, D_n),
$$ 
where $v$ summarizes agent-specific costs and benefits that are independent of prediction error, while $L^{*}(\pi, D_n)$ is the component that depends on prediction error---in line with our notation, the subjective MPSE.  The prior $\pi$ summarizes the agent's prior belief about the relationship between the unknown variable they need to predict (e.g., market demand for an entrepreneur, price movement for an investor) and various observables they consider relevant. The agent's model is the set of variables they consider relevant for prediction. The data $D_n$ is past data about this relationship. The agent knows $v$ and $\pi$, observes $D_n$, and then chooses the risky option if its expected utility is positive.

The results of this paper directly apply. \textit{Ceteris paribus} (fixing $v$), with few data points, agents with ``simple models'' are  systematically more confident in their prediction error ( $L^{*}(\pi, D_n)$ is lower) and are, therefore, more likely to take the risky option. Crucially, this holds whether their simple models are correct or not. This has an immediate implication that provides a novel comparative static: entrepreneurs with simple models are over-represented in sectors where little data has accumulated, even when the true DGP is complex. For example, this margin of selection suggests that, \textit{ceteris paribus}, the entrepreneurs more eager to invest in a country that just opened up to foreign investment, or in new technologies, will tend to be those that believe they can predict future conditions using relatively few covariates---that have a simple model---even when reality is much more complex. Similarly, investors with simpler models are more likely to enter into new asset classes (e.g., crypto-currencies). 

So far we have assumed that no new data is revealed after the investment decision. In reality, however, new data accumulates after the initial investment/ choice to enter is made.  Entrepreneurs or investors may take this into account in their decision, expecting to be able to refine their prediction. The discussion of Section \ref{sec:smallnchangeprior} suggests that this only \textit{strengthens} the selection in favor of simple models. To illustrate, consider the setup above but suppose that entrepreneurs must decide whether or not to invest  \textit{before} any data is revealed, but knowing that some data will be revealed at a later stage. As we discussed in Section \ref{sec:smallnchangeprior}, agents with simple models are more confident about how much they will be able to learn from the yet-to-be-released data. In this case, entrepreneurs/ investors with overly simplistic models will \textit{always be over represented}. %This implies that, when investments are to be decided before much of the information is revealed---but knowing that it will be---there will be even more substantial over-representation of agents with overly-simplistic models. 

\paragraph{Connections to Overconfidence.} These findings connect to established empirical facts on overconfidence and entry. Several studies have shown that entrepreneurs are, by various measures, overconfident \citep{koellinger2007think, COOPER198897}. Similarly, a large body of evidence shows that (especially retail) investors are often overconfident about their knowledge and information \citep{odean1999investors, statman2006investor}. This is commonly attributed to either incorrect beliefs, with selection favoring individuals with overly optimistic priors, or post-decision bolstering. % (having entered, the entrepreneur ``bolsters'' their choice to make it seem more attractive than they originally believed). 

Our results suggest a novel margin of selection related to model complexity in relation to overconfidence: entry into areas with limited past data (e.g., ``new'' areas) are systematically biased towards entrepreneurs or investors with models that are ``simple.'' As we have seen, when the true DGP is complex, these individuals are also \textit{overconfident} in their predictive ability: their subjective MPSE is on average lower than it should be. The relevant margin of selection may be the simplicity of the model, which generates both a higher likelihood of entrance and overconfidence in predictive ability. Similarly, investment in new asset classes is more common for investors who, \textit{ceteris paribus}, have simple predictive models of price movement. Our results also suggest that this margin of selection is attenuated as data accumulates: entry into more established sectors/ technologies/ countries may be systematically different from entry into ``new'' areas in terms of complexity of the entrepreneur's model; investors in established assets/ areas may be systematically different from investors in new asset classes/ trends (e.g., crypto-currency). 

Existing studies on overconfidence note also that agents often misreact, or underreact, to new information. For example, \cite{odean1999investors} studies the trading patterns of retail traders, and argues that their trades are systematically incorrect: ``while investors' overconfidence in the precision of their information may contribute to this finding, it is not sufficient to explain it. These investors must be systematically misinterpreting information available to them. They do not simply misconstrue the precision of their information, but its very meaning.'' This is consistent with our finding that, when the true DGP is complex and involves many variables, entry into trading may favor agents with incorrect models, i.e., ones that exclude covariates that are relevant for prediction and/or include irrelevant ones.

\subsection{Competing Factor Models\protect\footnote{We thank Stefano Giglio for useful discussions in writing this section.}} \label{sec:factorzoo}
A large body of work in finance has studied the cross-sectional variation in asset expected returns (i.e., why different assets earn different average returns). The classical asset-pricing framework \citep{jensen1972capital,fama1973risk} posits that, at each point in time, asset returns  are governed by a \emph{multi-factor} model. The  return of each asset---an individual stock or a portfolio---is an asset-specific linear combination of these factors  (with time-invariant coefficients) plus random noise. 

The search for factors to explain the cross-sectional variation of expected stock returns has produced hundreds of potential candidates. The literature has evolved from the parsimonious model of \cite{fama1993common} using only three factors (market return, size premium, value premium) to the factor library in \cite{feng2020taming} that contains 150 risk factors.\footnote{See Appendix \ref{subsection:additional_figures} for a description of these factors, how they are constructed, and the year in which they were published.}

We use our framework to understand the proliferation of factors in this literature. We argue that the increase in the number of test portfolios used to compute the Fama-French cross-sectional regressions mechanically favors asset-pricing models with several factors. 

To make this point, we view different collections of factors as competing models or, more precisely, competing sets of risk factors---a terminology that has been used, incidentally, by \cite{fama1993common, fama2015five}.%
\footnote{From \citet[p.\ 12]{fama1993common}: ``The average excess returns on the portfolios that serve as dependent variables give perspective on the range of average returns that competing sets of risk factors must explain.'' On p.\ 13: ``The wide range of average returns on the 25 stock portfolios, and the size and book-to-market effects in average returns, present interesting challenges for competing sets of risk factors.'' Finally, in \cite{fama2015five} ``estimate the proportion of the cross-section of expected returns left unexplained by competing models.'' } 
We then take the number of test portfolios as the number of available data points to predict the cross-section of expected returns. We consider the different factor models as different Bayesian agents competing to predict cross-sectional returns. %
%\footnote{As an aside, we note that \cite{harvey2017presidential} in his 2017 AFA presidential address called for Bayesian solutions to the factor zoo.} 
Our results suggest that the winning model depends crucially on the sample size. With a few test portfolios, smaller factor models will be selected. %
%\footnote{Our results are therefore consistent with the recent work of \cite{bryzgalova2019bayesian} who conclude that a three most likely factor model outperforms most benchmarks on a set of $60$ test portfolios and $51$ potential factors. However we note that their methods, while Bayesian, are quite different to ours.}
Conversely, increasing the number of test portfolios favors high-dimensional factor models.

Let $i$ index an asset in the cross-section. The outcome variable $Y_i$ denotes the excess return of asset $i$ averaged over the different time periods for which data on returns and factors are available.\footnote{We follow \cite{feng2020taming} and use monthly returns from July 1976 to December 2017.} Each asset $i$ has an associated 150-dimensional vector of covariates, $X_i$, containing the asset's factor loadings.\footnote{In the standard Fama-French two-pass regressions, these factor loadings are estimated from asset-by-asset time series regressions of excess returns on factors \citep{BAI201531}. For simplicity, we ignore the estimation error and treat the estimated factor loadings as the true factor loadings. This allows us to stay within our simple linear regression framework. The competing models are the different subsets of the factors that different agents believe to be relevant to explain the cross-section of asset expected returns.} In principle, assets could be individual stocks or portfolios. We follow the literature and focus on portfolios.\footnote{There is some discussion in the literature about what is the right unit of observation to test asset-pricing models: see the discussion in   \cite{ang_liu_schwarz_2020}.} The number of portfolios under consideration $(N)$ gives the sample size for the cross-sectional regressions. 

We start with the 25 (5 $\times$ 5) portfolios sorted by size and book-to-market ratio as used in \cite{fama1993common}.\footnote{A 5 $\times$ 5 bivariate portfolio refers to the common practice of grouping stocks by the quintiles of the cross-sectional distribution of both size and book-to-market.} We then consider the larger collection of 2,875 (5 $\times$ 5) bivariate-sorted portfolios in \cite{feng2020taming}.\footnote{The sorting is based on size and each of the 115 factors marked with asterix in the table in Appendix \ref{app:c3}. The dataset was obtained directly from the replication files provided by \cite{feng2020taming}.}

\paragraph{Competing (Factor) Models.} We consider ``competition'' between three different models. The first (Fama-French) uses only the original Fama-French factors (excess market return, and the ``small minus big'' and ``high minus low'' factors). The second model (Factor Zoo) uses all 150 factors in \cite{feng2020taming}. The third (FGX-DS) uses the 135 factors introduced up until 2011, plus five additional factors obtained by the ``Double Selection'' approach of \cite{feng2020taming}.\footnote{These are the  investment and profitability factors of \cite{hou2015digesting}, the ``robust minus weak'' factor of \cite{fama2015five}, the intermediary risk factor of \cite{he2017intermediary}, and the ``Quality Minus Junk'' factor of \cite{asness2019quality}.}  For each model, we consider the same set of hyper-parameters $(a_0,b_0)$, chosen to maximize the marginal likelihood of the largest model with the largest dataset, and set $\gamma$ so that all models have the same prior subjective MPSE before any data. Details are provided in Appendix \ref{section:Priors-hyper}.%\footnote{The marginal likelihood refers to the density of $Y|X,a_0,b_0,\gamma$ after $\beta$ and $\sigma^2$ have been integrated out.}  

\begin{figure}[t!]
\centering
\includegraphics[keepaspectratio, scale=.5]{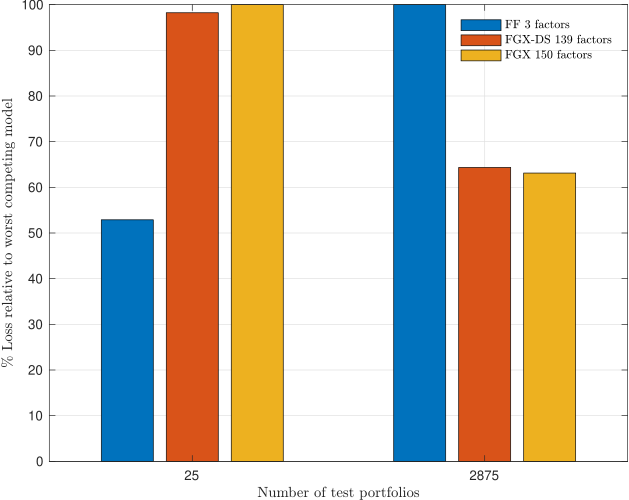}\\
\caption{Competing (Factor) Models}
\label{fig:CFM}
\end{figure}

Figure \ref{fig:CFM} presents our results. It shows the subjective MSPE of each model. To make units easier to interpret, the results are presented as a percentage relative to the worst competing model (out of the three under consideration).  

Consistent with our theorems, if we consider only the 25 ($5\, \times\, 5$) bivariate sorted portfolios on size and book-to-market (as \cite{fama1993common} originally did), the simple three-factor model achieves the best subjective MPSE (around half of the subjective MPSE obtained with the largest models). With 2,875 ($5 \times 5$) portfolios, however, the ranking reverses, and larger models now have the lowest subjective MPSE. (Appendix \ref{subsection:additional_figures} presents several robustness checks with different models,  sample sizes, and subsets of portfolios.)

\paragraph{A General Model of Scientific Progress.} This discussion suggests a simple model of scientific progress. There is public interest in predicting a variable $y$ as a function of observed covariates $x$, and there are competing scientists, each described by a prior belief about the world; these priors differ in what covariates they think are relevant for predicting $y$. Data (publicly) accumulates as i.i.d.\ draws from the unknown DGP.

Suppose each model's success also depends on its subjective MSPE. Scientists who believe their model has low prediction error would be more forceful about it, staking their career on its predictions. Others whose subjective MSPE is high may be worried about mistakes and damage to their reputation. Practitioners or politicians, who may cite scientific research to justify their actions, may be more prone to adopt models with low subjective MSPE.

With these assumptions, our results suggest the following dynamic of scientific progress. In the early stages of a field, when data is relatively scarce, overly simple models prevail---including frameworks that exclude relevant covariates. Over time, more and more data accumulates, and more nuanced models come into vogue, involving ever-increasing collections of covariates. Overly-simple models are then discarded, since they are unable to fit the data as well as larger ones; the ``scientific paradigm,'' understood as the collection of relevant variables, becomes more complex. This is in line with casual observation and with dynamics described in epistemology. For example, this aligns with what \cite{kuhn2012structure} describes as the path of progress of ``normal'' science, i.e., after a dominant paradigm has been established.\footnote{Changing paradigms are outside the scope of this work---see, e.g., \cite{ortoleva2012modeling} for a model of a non-Bayesian decision-maker who changes paradigm (selects a new prior) upon receiving information that is unexpected according to their current prior.}

\section{Related Literature} \label{sec:literature}

A large body of literature has studied model misspecification in individual decision-making, with famous examples like overconfidence and correlation neglect. A few recent theoretical contributions to this enormous literature include \cite{heidhues2018unrealistic} and \cite{ortoleva2015overconfidence}, to which we refer for further references. In misspecified learning settings, ``feedback loops'' between the agents' misspecified beliefs and the action they take add further technical challenges---see, e.g., \cite{fudenberg2017active}, \cite{fudenberg2020limits}, \cite{heidhues2020convergence}. 

Recent works have studied the implications of agents with misspecified models in various strategic settings. For instance,  \cite{bohren2016informational}, \cite{bohren2017bounded},  \cite{frick2019misinterpreting}, and \cite{frick2019dispersed} study social learning when agents have misspecified models that cause them to misinterpret other agents' actions. \cite{mailath2019wisdom} study a stylized prediction market where Bayesian agents have different models (defined as different partitions of a common state space) and discuss the possibility of information aggregation. 

Recent works consider specifically the outcomes when agents' models are misspecified in the sense we study here, i.e., there is a payoff-relevant dependent variable, and agents either include irrelevant independent variables or exclude dependent variables. \cite{schwartzstein2021using} considers this in the context of persuasion, where competing persuaders may ``overfit'' the data to better persuade a receiver. \cite{levy2019misspecified} study a political economy setting where there are both ``simple'' world views and complicated ones, and ask whether political competition disciplines overly simplistic world views, finding instead that they recur in dynamic settings. Finally, \cite{he2020evolutionarily} ask whether (and what kind of) misspecifications can be evolutionarily stable. These works find reasons for why misspecified models may survive (overfit models in the former, simple models in the case of the latter two), although the exact mechanism is different than ours.  Recent work also considers the possibility that agents with misspecified models may be able to realize it, and  characterize the kinds of misspecifications that survive---see, e.g., \cite{fudenberg2020misperceptions}, who consider an evolutionary framework, or \cite{gagnon2021channeled}, who study a setting where the agent perceives their errors through the framework of their own model.

In strategic settings, \cite{esponda2016berk} define a learning-based solution concept (``Berk-Nash Equilibrium'') for games in which agents' beliefs are misspecified. More broadly, solution concepts have been posited for settings where agents suffer from some sort of misspecification, including well-known examples like analogy-based equilibrium \citep{jehiel2005analogy} and cursed equilibrium \citep{eyster2005cursed}.

Several works consider outcomes when some agents behave in a way that can be construed as coming from a misspecified model. For instance, in \cite{spiegler2006market, spiegler2013placebo}, society misunderstands the relationship between outcomes and the actions of strategic agents, which affects the actions these agents take in equilibrium and resulting outcomes (studied in the context of a market for quacks or its implications for political reforms).  \cite{levy2019misspecified} study a dynamic model of political competition where agents have different (misspecified) models of the world; the study uses this model to provide a foundation for the recurrence of populism.  
\cite{liang2018games} studies outcomes in games of incomplete information where agents behave like statisticians and have limited information.\footnote{There is a larger literature that studies the outcomes when agents are modeled as statisticians or machine learners, e.g., \cite{al2009decision}, \cite{al2014coarse}, \cite{acemoglu2016fragility} and \cite{cherry2006statistical}.}

A novel approach to modeling misspecification in economic theory is the directed acyclic graph approach; see \cite{pearl2009causality}. This is exploited in a single-person decision framework in \cite{spiegler2016bayesian}, which studies a single decision maker with a misspecified causal model and large amounts of data. The paper shows that the decision maker may evaluate actions differently than their long-run frequencies, and exhibit artifacts such as ``reverse causation'' and coarse decision making. This approach is then used in \cite{eliaz2018model}, which proposes a model of competing narratives. A narrative is a causal model that maps actions into consequences, including other random, unrelated variables. An equilibrium notion is defined, and the paper studies the distribution of narratives that is obtained in equilibrium.

Finally, the understanding that agents should be cognizant that their models may be misspecified has also led to new approaches in mechanism design, where the designer accounts for misspecification in various ways.  The literature on robust mechanism design (beginning with the seminal \citealt{bergemann2005robust}) provides foundations for using stronger solution concepts.  \cite{madarasz2017sellers} show that an optimal mechanism may perform very poorly if the planner's  model is even slightly misspecified, and they identify a class of near optimal mechanisms that degrade gracefully. Works such as \cite{chassang2013calibrated} and  \cite{carroll2015robustness} develop optimal ``robust'' contracts and contrast to classical optimal contracting. 

Since one natural application of our model is an auction, our results are related to \cite{atakan2014auctions}, who consider the competitive sale of assets whose value depends on how they are utilized.\footnote{\cite{bond2010information} study a trading environment with a similar feature.} The successful bidder chooses an action that determines, together with the state of the world, the payoff generated by the asset. They focus on a setting where  bidders have a common prior but observe private signals. Their main result is the possibility of (complete) failure of information aggregation. Our model is similar in that the value of the object depends on an action taken by the agent. However, our work considers a complementary environment where all bidders observe the same information but  have different priors. Information aggregation is ruled out by assumption, and our key theme is model selection.

We assume that agents have different priors and are fully aware they have different priors: that is to say, our agents agree to disagree. This assumption has been used in economic theory at least since \cite{harrison1978speculative}. We refer the reader to \cite{morris1995common} for a discussion of the common and heterogeneous prior traditions in economic theory. Heterogenous priors have been used in a number of applications in bargaining \citep{yildiz2003bargaining}, trade \citep{morris1994trade},  financial markets \citep{scheinkman2003overconfidence,ottaviani2015price}, and more.  

%This provides an example of a set-up in which people adopt distorted beliefs to enhance their anticipatory
%utility.

\paragraph{Relation to Model Selection.}
Large literatures in statistics, econometrics, and machine learning study model selection methods and provide normative foundations; see \cite{ModelSelection08} and \cite{burnham2003model} for textbook overviews. Popular approaches include, for example, the $C_p$ criterion of  \cite{Mallows}, the Akaike Information Criterion (AIC) of \cite{akaike1974new}, and the Bayes Information Criterion (BIC) of \cite{schwarz1978}. 
We showed that is a connection between our large data results and the AIC introduced in \cite{akaike1974new}, in particular, to the asymptotic properties of the AIC characterized in the seminal paper of \cite{nishii1984asymptotic}.

While some of our asymptotic results are reminiscent of the model selection literature, there are three important differences. First, the aims of this literature are very different from ours. Ours is a positive approach of studying which model emerges from a competition between Bayesian agents with misspecified models. The approach in the model selection literature is instead \textit{normative}: various methods of model selection are proposed and studied with a view to avoiding over-fitting and/or selecting ``good'' models according to some metric. The results we are aware of speak to the asymptotic efficiency of these techniques. Second, not only are our results derived from a completely different setting, but they are also proven with different techniques.  Third, the connection is limited to the large-data result. We are not aware of any analogs to our small-sample results.

% We also use techniques and approaches from the statistics and econometrics literature.  The proof of Theorem \ref{prop:winner-fixed-n} uses `non-standard asymptotics' that allow for the parameters of a statistical model to be indexed by the sample size have been used extensively in econometrics. The typical goal of an alternative asymptotic framework is to provide better approximations to finite-sample distributions of estimators, tests, and confidence intervals, while exploiting Laws of Large Numbers and Central Limit Theorems. For example, the local-to-unity asymptotics of \cite{Phillips87} studies auto-regressive models that are close to being nonstationary; the local-to-zero asymptotics of \cite{SS97} studies Instrumental Variables models that are close to being unidentified; and \cite{Cattaneo2018} studies models where possibly many covariates are included for estimation and inference.

\section{Discussion and Conclusion} \label{sec:conclude}

A variable of interest is related to a vector of covariates. Different agents have different models of this relationship: in particular they rule in/rule out different covariates as being potentially related to prediction. All agents observe a common dataset of size $n$, drawn from the true DGP. We ask: Who is the agent with the highest confidence in their own predictive ability, in the form of the lowest mean-squared prediction error according to their own subjective posterior?  We study the relationship between sample size and the dimension of the winning model. This applies to all cases in which confidence in predictive ability affects selection. We show results of two kinds.

First, when $n$ is small, models that employ few covariates may take the lead, even if the true DGP is more complex. To establish this result formally, we use Normal-Inverse-Gamma-Inverse-Wishart priors. 
 Second, when $n$ is large, misspecified models (i.e., models that rule out an observable that is relevant for prediction) never win. However, high-dimensional models that include irrelevant covariates (but do not exclude relevant ones) may continue to win. Our results show that the effect of the prior on model competition does not vanish in large samples. These results hold for a very general class of priors and true DGPs. 

Finally, we give two applications. First, we apply our results to a model of entry: entrepreneurs decide whether to enter a new market, households decide whether to invest in a new asset class. We show that, insofar as prediction error of future variables is relevant for profitability, our results suggest a new margin of selection: when data is relatively scarce, agents with simpler models will be over-represented in the entry decision. Our second application is to understand the proliferation of factors that explain the cross-sectional variation of expected stock returns in the asset-pricing literature. We show how the increase in the number of test portfolios used to compute the cross-sectional regressions mechanically favors models with several factors. 

% There are several natural avenues to future research. An obvious one is a setting in which agents each observe a private dataset: this complicates our analysis because now a notion of the winner's curse applies. Another one is to consider dynamic variants: if agents got feedback or could invest to acquire more data, what kinds of models would be selected?

\newpage

\bibliographystyle{ecta}
\bibliography{sample}

\begin{thebibliography}{74}
\newcommand{\enquote}[1]{``#1''}
\expandafter\ifx\csname natexlab\endcsname\relax\def\natexlab#1{#1}\fi

\bibitem[\protect\citeauthoryear{Acemoglu, Chernozhukov, and Yildiz}{Acemoglu
  et~al.}{2016}]{acemoglu2016fragility}
\textsc{Acemoglu, D., V.~Chernozhukov, and M.~Yildiz} (2016):
  \enquote{Fragility of asymptotic agreement under Bayesian learning,}
  \emph{Theoretical Economics}, 11, 187--225.

\bibitem[\protect\citeauthoryear{Akaike}{Akaike}{1974}]{akaike1974new}
\textsc{Akaike, H.} (1974): \enquote{A new look at the statistical model
  identification,} \emph{IEEE transactions on automatic control}, 19, 716--723.

\bibitem[\protect\citeauthoryear{Al-Najjar}{Al-Najjar}{2009}]{al2009decision}
\textsc{Al-Najjar, N.~I.} (2009): \enquote{Decision makers as statisticians:
  Diversity, ambiguity, and learning,} \emph{Econometrica}, 77, 1371--1401.

\bibitem[\protect\citeauthoryear{Al-Najjar and Pai}{Al-Najjar and
  Pai}{2014}]{al2014coarse}
\textsc{Al-Najjar, N.~I. and M.~M. Pai} (2014): \enquote{Coarse decision making
  and overfitting,} \emph{Journal of Economic Theory}, 150, 467--486.

\bibitem[\protect\citeauthoryear{Alonso, Dessein, and Matouschek}{Alonso
  et~al.}{2008}]{alonso2008does}
\textsc{Alonso, R., W.~Dessein, and N.~Matouschek} (2008): \enquote{When does
  coordination require centralization?} \emph{American Economic Review}, 98,
  145--79.

\bibitem[\protect\citeauthoryear{Ang, Liu, and Schwarz}{Ang
  et~al.}{2020}]{ang_liu_schwarz_2020}
\textsc{Ang, A., J.~Liu, and K.~Schwarz} (2020): \enquote{Using Stocks or
  Portfolios in Tests of Factor Models,} \emph{Journal of Financial and
  Quantitative Analysis}, 55, 709–750.

\bibitem[\protect\citeauthoryear{Asness, Frazzini, and Pedersen}{Asness
  et~al.}{2019}]{asness2019quality}
\textsc{Asness, C.~S., A.~Frazzini, and L.~H. Pedersen} (2019):
  \enquote{Quality minus junk,} \emph{Review of Accounting Studies}, 24,
  34--112.

\bibitem[\protect\citeauthoryear{Atakan and Ekmekci}{Atakan and
  Ekmekci}{2014}]{atakan2014auctions}
\textsc{Atakan, A.~E. and M.~Ekmekci} (2014): \enquote{Auctions, actions, and
  the failure of information aggregation,} \emph{American Economic Review},
  104.

\bibitem[\protect\citeauthoryear{Bai and Zhou}{Bai and Zhou}{2015}]{BAI201531}
\textsc{Bai, J. and G.~Zhou} (2015): \enquote{Fama–MacBeth two-pass
  regressions: Improving risk premia estimates,} \emph{Finance Research
  Letters}, 15, 31--40.

\bibitem[\protect\citeauthoryear{Bergemann and Morris}{Bergemann and
  Morris}{2005}]{bergemann2005robust}
\textsc{Bergemann, D. and S.~Morris} (2005): \enquote{Robust mechanism design,}
  \emph{Econometrica}, 73, 1771--1813.

\bibitem[\protect\citeauthoryear{Berk}{Berk}{1970}]{berk1970consistency}
\textsc{Berk, R.~H.} (1970): \enquote{Consistency a posteriori,} \emph{The
  Annals of Mathematical Statistics}, 894--906.

\bibitem[\protect\citeauthoryear{Bishop}{Bishop}{2006}]{bishop2006pattern}
\textsc{Bishop, C.~M.} (2006): \emph{Pattern recognition and machine learning},
  springer.

\bibitem[\protect\citeauthoryear{Bohren}{Bohren}{2016}]{bohren2016informational}
\textsc{Bohren, J.~A.} (2016): \enquote{Informational herding with model
  misspecification,} \emph{Journal of Economic Theory}, 163, 222--247.

\bibitem[\protect\citeauthoryear{Bohren and Hauser}{Bohren and
  Hauser}{2017}]{bohren2017bounded}
\textsc{Bohren, J.~A. and D.~Hauser} (2017): \enquote{Bounded rationality and
  learning: A framework and a robustness result,} \emph{Working Paper,
  University of Pennsylvania}.

\bibitem[\protect\citeauthoryear{Bond and Eraslan}{Bond and
  Eraslan}{2010}]{bond2010information}
\textsc{Bond, P. and H.~Eraslan} (2010): \enquote{Information-based trade,}
  \emph{Journal of Economic Theory}, 145, 1675--1703.

\bibitem[\protect\citeauthoryear{Burnham and Anderson}{Burnham and
  Anderson}{2003}]{burnham2003model}
\textsc{Burnham, K.~P. and D.~R. Anderson} (2003): \emph{Model selection and
  multimodel inference: a practical information-theoretic approach}, Springer
  Science \& Business Media.

\bibitem[\protect\citeauthoryear{Carroll}{Carroll}{2015}]{carroll2015robustness}
\textsc{Carroll, G.} (2015): \enquote{Robustness and linear contracts,}
  \emph{American Economic Review}, 105, 536--63.

\bibitem[\protect\citeauthoryear{Chassang}{Chassang}{2013}]{chassang2013calibrated}
\textsc{Chassang, S.} (2013): \enquote{Calibrated incentive contracts,}
  \emph{Econometrica}, 81, 1935--1971.

\bibitem[\protect\citeauthoryear{Cherry and Salant}{Cherry and
  Salant}{2018}]{cherry2006statistical}
\textsc{Cherry, J. and Y.~Salant} (2018): \enquote{Statistical Inference in
  Games,} Tech. rep., mimeo.

\bibitem[\protect\citeauthoryear{Claeskens and Hjort}{Claeskens and
  Hjort}{2008}]{ModelSelection08}
\textsc{Claeskens, G. and N.~Hjort} (2008): \enquote{Model selection and model
  averaging,} \emph{Cambridge Books}.

\bibitem[\protect\citeauthoryear{Cooper, Woo, and Dunkelberg}{Cooper
  et~al.}{1988}]{COOPER198897}
\textsc{Cooper, A.~C., C.~Y. Woo, and W.~C. Dunkelberg} (1988):
  \enquote{Entrepreneurs' perceived chances for success,} \emph{Journal of
  Business Venturing}, 3, 97--108.

\bibitem[\protect\citeauthoryear{Eliaz and Spiegler}{Eliaz and
  Spiegler}{2018}]{eliaz2018model}
\textsc{Eliaz, K. and R.~Spiegler} (2018): \enquote{A Model of Competing
  Narratives,} \emph{CEPR Discussion Paper No. DP13319}.

\bibitem[\protect\citeauthoryear{Esponda and Pouzo}{Esponda and
  Pouzo}{2016}]{esponda2016berk}
\textsc{Esponda, I. and D.~Pouzo} (2016): \enquote{Berk--Nash equilibrium: A
  framework for modeling agents with misspecified models,} \emph{Econometrica},
  84, 1093--1130.

\bibitem[\protect\citeauthoryear{Eyster and Rabin}{Eyster and
  Rabin}{2005}]{eyster2005cursed}
\textsc{Eyster, E. and M.~Rabin} (2005): \enquote{Cursed equilibrium,}
  \emph{Econometrica}, 73, 1623--1672.

\bibitem[\protect\citeauthoryear{Fama and French}{Fama and
  French}{1993}]{fama1993common}
\textsc{Fama, E.~F. and K.~R. French} (1993): \enquote{Common risk factors in
  the returns on stocks and bonds,} \emph{Journal of financial economics}, 33,
  3--56.

\bibitem[\protect\citeauthoryear{Fama and French}{Fama and
  French}{2015}]{fama2015five}
---\hspace{-.1pt}---\hspace{-.1pt}--- (2015): \enquote{A five-factor asset
  pricing model,} \emph{Journal of financial economics}, 116, 1--22.

\bibitem[\protect\citeauthoryear{Fama and MacBeth}{Fama and
  MacBeth}{1973}]{fama1973risk}
\textsc{Fama, E.~F. and J.~D. MacBeth} (1973): \enquote{Risk, return, and
  equilibrium: Empirical tests,} \emph{Journal of political economy}, 81,
  607--636.

\bibitem[\protect\citeauthoryear{Feng, Giglio, and Xiu}{Feng
  et~al.}{2020}]{feng2020taming}
\textsc{Feng, G., S.~Giglio, and D.~Xiu} (2020): \enquote{Taming the factor
  zoo: A test of new factors,} \emph{The Journal of Finance}, 75, 1327--1370.

\bibitem[\protect\citeauthoryear{Frick, Iijima, and Ishii}{Frick
  et~al.}{2019{\natexlab{a}}}]{frick2019dispersed}
\textsc{Frick, M., R.~Iijima, and Y.~Ishii} (2019{\natexlab{a}}):
  \enquote{Dispersed Behavior and Perceptions in Assortative Societies,} .

\bibitem[\protect\citeauthoryear{Frick, Iijima, and Ishii}{Frick
  et~al.}{2019{\natexlab{b}}}]{frick2019misinterpreting}
---\hspace{-.1pt}---\hspace{-.1pt}--- (2019{\natexlab{b}}):
  \enquote{Misinterpreting Others and the Fragility of Social Learning,}
  \emph{Cowles Foundation Discussion Paper}.

\bibitem[\protect\citeauthoryear{Fudenberg and Lanzani}{Fudenberg and
  Lanzani}{2020}]{fudenberg2020misperceptions}
\textsc{Fudenberg, D. and G.~Lanzani} (2020): \enquote{Which misperceptions
  persist?} \emph{Available at SSRN}.

\bibitem[\protect\citeauthoryear{Fudenberg, Lanzani, and Strack}{Fudenberg
  et~al.}{2020}]{fudenberg2020limits}
\textsc{Fudenberg, D., G.~Lanzani, and P.~Strack} (2020): \enquote{Limits
  Points of Endogenous Misspecified Learning,} \emph{Available at SSRN}.

\bibitem[\protect\citeauthoryear{Fudenberg, Romanyuk, and Strack}{Fudenberg
  et~al.}{2017}]{fudenberg2017active}
\textsc{Fudenberg, D., G.~Romanyuk, and P.~Strack} (2017): \enquote{Active
  learning with a misspecified prior,} \emph{Theoretical Economics}, 12,
  1155--1189.

\bibitem[\protect\citeauthoryear{Gagnon-Bartsch, Rabin, and
  Schwartzstein}{Gagnon-Bartsch et~al.}{2021}]{gagnon2021channeled}
\textsc{Gagnon-Bartsch, T., M.~Rabin, and J.~Schwartzstein} (2021):
  \enquote{Channeled Attention and Stable Errors,} \emph{Working Paper}.

\bibitem[\protect\citeauthoryear{Greene}{Greene}{2018}]{greene2018econometric}
\textsc{Greene, W.~H.} (2018): \emph{Econometric Analysis}, vol. 8th Edition,
  Pearson.

\bibitem[\protect\citeauthoryear{Hansen}{Hansen}{2021}]{hansen2020econometrics}
\textsc{Hansen, B.} (2021): \enquote{Econometrics,} \emph{A textbook draft
  available online at www. ssc. wisc. edu/\~{}
  bhansen/econometrics/Econometrics. pdf}.

\bibitem[\protect\citeauthoryear{Harrison and Kreps}{Harrison and
  Kreps}{1978}]{harrison1978speculative}
\textsc{Harrison, J.~M. and D.~M. Kreps} (1978): \enquote{Speculative investor
  behavior in a stock market with heterogeneous expectations,} \emph{The
  Quarterly Journal of Economics}, 92, 323--336.

\bibitem[\protect\citeauthoryear{He and Libgober}{He and
  Libgober}{2020}]{he2020evolutionarily}
\textsc{He, K. and J.~Libgober} (2020): \enquote{Evolutionarily Stable (Mis)
  specifications: Theory and Applications,} \emph{arXiv preprint
  arXiv:2012.15007}.

\bibitem[\protect\citeauthoryear{He, Kelly, and Manela}{He
  et~al.}{2017}]{he2017intermediary}
\textsc{He, Z., B.~Kelly, and A.~Manela} (2017): \enquote{Intermediary asset
  pricing: New evidence from many asset classes,} \emph{Journal of Financial
  Economics}, 126, 1--35.

\bibitem[\protect\citeauthoryear{Heidhues, K{\H{o}}szegi, and Strack}{Heidhues
  et~al.}{2018}]{heidhues2018unrealistic}
\textsc{Heidhues, P., B.~K{\H{o}}szegi, and P.~Strack} (2018):
  \enquote{Unrealistic expectations and misguided learning,}
  \emph{Econometrica}, 86, 1159--1214.

\bibitem[\protect\citeauthoryear{Heidhues, Koszegi, and Strack}{Heidhues
  et~al.}{2020}]{heidhues2020convergence}
\textsc{Heidhues, P., B.~Koszegi, and P.~Strack} (2020): \enquote{Convergence
  in Models of Misspecified Learning,} .

\bibitem[\protect\citeauthoryear{Hogg, Mckean, and Allen}{Hogg
  et~al.}{2006}]{hogg}
\textsc{Hogg, R.~V., J.~W. Mckean, and C.~D. Allen} (2006): \emph{Introduction
  To Mathematical Statistics}, Pearson Education India.

\bibitem[\protect\citeauthoryear{Hou, Xue, and Zhang}{Hou
  et~al.}{2015}]{hou2015digesting}
\textsc{Hou, K., C.~Xue, and L.~Zhang} (2015): \enquote{Digesting anomalies: An
  investment approach,} \emph{The Review of Financial Studies}, 28, 650--705.

\bibitem[\protect\citeauthoryear{Jehiel}{Jehiel}{2005}]{jehiel2005analogy}
\textsc{Jehiel, P.} (2005): \enquote{Analogy-based expectation equilibrium,}
  \emph{Journal of Economic theory}, 123, 81--104.

\bibitem[\protect\citeauthoryear{Jensen, Black, and Scholes}{Jensen
  et~al.}{1972}]{jensen1972capital}
\textsc{Jensen, M.~C., F.~Black, and M.~S. Scholes} (1972): \enquote{The
  capital asset pricing model: Some empirical tests,} .

\bibitem[\protect\citeauthoryear{Kass, Tierney, and Kadane}{Kass
  et~al.}{1990}]{KTK:1990}
\textsc{Kass, R., L.~Tierney, and J.~B. Kadane} (1990): \enquote{The validity
  of posterior expansions based on Laplaces method,} in \emph{Bayesian and
  Likelihood Methods in Statistics and Econometrics}, ed. by S.~Geisser,
  J.~Hodges, S.~Press, and A.~Zellner, vol.~7, 473.

\bibitem[\protect\citeauthoryear{Kleijn and Van~der Vaart}{Kleijn and Van~der
  Vaart}{2012}]{kleijn2012bernstein}
\textsc{Kleijn, B. and A.~Van~der Vaart} (2012): \enquote{The
  Bernstein-von-Mises theorem under misspecification,} \emph{Electronic Journal
  of Statistics}, 6, 354--381.

\bibitem[\protect\citeauthoryear{Koellinger, Minniti, and Schade}{Koellinger
  et~al.}{2007}]{koellinger2007think}
\textsc{Koellinger, P., M.~Minniti, and C.~Schade} (2007): \enquote{“I think
  I can, I think I can”: Overconfidence and entrepreneurial behavior,}
  \emph{Journal of economic psychology}, 28, 502--527.

\bibitem[\protect\citeauthoryear{Kozak, Nagel, and Santosh}{Kozak
  et~al.}{2018}]{kozak2018interpreting}
\textsc{Kozak, S., S.~Nagel, and S.~Santosh} (2018): \enquote{{Interpreting
  Factor Models},} \emph{Journal of Finance}, 73, 1183--1223.

\bibitem[\protect\citeauthoryear{Kuhn}{Kuhn}{1962}]{kuhn2012structure}
\textsc{Kuhn, T.~S.} (1962): \emph{The structure of scientific revolutions},
  University of Chicago press.

\bibitem[\protect\citeauthoryear{Levy, Razin, and Young}{Levy
  et~al.}{2019}]{levy2019misspecified}
\textsc{Levy, G., R.~Razin, and A.~Young} (2019): \enquote{Misspecified
  Politics and the Recurrence of Populism,} Tech. rep., Working Paper.

\bibitem[\protect\citeauthoryear{Liang}{Liang}{2018}]{liang2018games}
\textsc{Liang, A.} (2018): \enquote{Games of Incomplete Information Played by
  Statisticians,} \emph{Working paper, University of Pennsylvania}.

\bibitem[\protect\citeauthoryear{Madar{\'a}sz and Prat}{Madar{\'a}sz and
  Prat}{2017}]{madarasz2017sellers}
\textsc{Madar{\'a}sz, K. and A.~Prat} (2017): \enquote{Sellers with
  misspecified models,} \emph{The Review of Economic Studies}, 84, 790--815.

\bibitem[\protect\citeauthoryear{Mailath and Samuelson}{Mailath and
  Samuelson}{2019}]{mailath2019wisdom}
\textsc{Mailath, G.~J. and L.~Samuelson} (2019): \enquote{The Wisdom of a
  Confused Crowd: Model-Based Inference,} \emph{Cowles Foundation Discussion
  Paper}.

\bibitem[\protect\citeauthoryear{Mallows}{Mallows}{1973}]{Mallows}
\textsc{Mallows, C.~L.} (1973): \enquote{Some Comments on C<sub>P</sub>,}
  \emph{Technometrics}, 15, 661--675.

\bibitem[\protect\citeauthoryear{Marschak and Radner}{Marschak and
  Radner}{1972}]{marschak1972economic}
\textsc{Marschak, J. and R.~Radner} (1972): \emph{Economic Theory of Teams.}

\bibitem[\protect\citeauthoryear{McCullagh}{McCullagh}{2002}]{mccullagh2002statistical}
\textsc{McCullagh, P.} (2002): \enquote{What is a statistical model?} \emph{The
  Annals of Statistics}, 30, 1225--1310.

\bibitem[\protect\citeauthoryear{Milgrom and Roberts}{Milgrom and
  Roberts}{1992}]{roberts1992economics}
\textsc{Milgrom, P. and J.~Roberts} (1992): \emph{Economics, organization and
  management}, Prentice-Hall Englewood Cliffs, NJ.

\bibitem[\protect\citeauthoryear{Morris}{Morris}{1994}]{morris1994trade}
\textsc{Morris, S.} (1994): \enquote{Trade with heterogeneous prior beliefs and
  asymmetric information,} \emph{Econometrica: Journal of the Econometric
  Society}, 1327--1347.

\bibitem[\protect\citeauthoryear{Morris}{Morris}{1995}]{morris1995common}
---\hspace{-.1pt}---\hspace{-.1pt}--- (1995): \enquote{The common prior
  assumption in economic theory,} \emph{Economics \& Philosophy}, 11, 227--253.

\bibitem[\protect\citeauthoryear{Nishii}{Nishii}{1984}]{nishii1984asymptotic}
\textsc{Nishii, R.} (1984): \enquote{Asymptotic properties of criteria for
  selection of variables in multiple regression,} \emph{The Annals of
  Statistics}, 758--765.

\bibitem[\protect\citeauthoryear{Odean}{Odean}{1999}]{odean1999investors}
\textsc{Odean, T.} (1999): \enquote{Do investors trade too much?}
  \emph{American economic review}, 89, 1279--1298.

\bibitem[\protect\citeauthoryear{Ortoleva}{Ortoleva}{2012}]{ortoleva2012modeling}
\textsc{Ortoleva, P.} (2012): \enquote{Modeling the change of paradigm:
  Non-Bayesian reactions to unexpected news,} \emph{American Economic Review},
  102, 2410--36.

\bibitem[\protect\citeauthoryear{Ortoleva and Snowberg}{Ortoleva and
  Snowberg}{2015}]{ortoleva2015overconfidence}
\textsc{Ortoleva, P. and E.~Snowberg} (2015): \enquote{Overconfidence in
  political behavior,} \emph{American Economic Review}, 105, 504--35.

\bibitem[\protect\citeauthoryear{Ottaviani and S{\o}rensen}{Ottaviani and
  S{\o}rensen}{2015}]{ottaviani2015price}
\textsc{Ottaviani, M. and P.~N. S{\o}rensen} (2015): \enquote{Price reaction to
  information with heterogeneous beliefs and wealth effects: Underreaction,
  momentum, and reversal,} \emph{American Economic Review}, 105, 1--34.

\bibitem[\protect\citeauthoryear{Pearl}{Pearl}{2009}]{pearl2009causality}
\textsc{Pearl, J.} (2009): \emph{Causality}, Cambridge university press.

\bibitem[\protect\citeauthoryear{Scheinkman and Xiong}{Scheinkman and
  Xiong}{2003}]{scheinkman2003overconfidence}
\textsc{Scheinkman, J.~A. and W.~Xiong} (2003): \enquote{Overconfidence and
  speculative bubbles,} \emph{Journal of political Economy}, 111, 1183--1220.

\bibitem[\protect\citeauthoryear{Schwartzstein and Sunderam}{Schwartzstein and
  Sunderam}{2021}]{schwartzstein2021using}
\textsc{Schwartzstein, J. and A.~Sunderam} (2021): \enquote{Using models to
  persuade,} \emph{American Economic Review}, 111, 276--323.

\bibitem[\protect\citeauthoryear{Schwarz}{Schwarz}{1978}]{schwarz1978}
\textsc{Schwarz, G.} (1978): \enquote{Estimating the Dimension of a Model,}
  \emph{Ann. Statist.}, 6, 461--464.

\bibitem[\protect\citeauthoryear{Spiegler}{Spiegler}{2006}]{spiegler2006market}
\textsc{Spiegler, R.} (2006): \enquote{The market for quacks,} \emph{The Review
  of Economic Studies}, 73, 1113--1131.

\bibitem[\protect\citeauthoryear{Spiegler}{Spiegler}{2013}]{spiegler2013placebo}
---\hspace{-.1pt}---\hspace{-.1pt}--- (2013): \enquote{Placebo reforms,}
  \emph{American Economic Review}, 103, 1490--1506.

\bibitem[\protect\citeauthoryear{Spiegler}{Spiegler}{2016}]{spiegler2016bayesian}
---\hspace{-.1pt}---\hspace{-.1pt}--- (2016): \enquote{Bayesian networks and
  boundedly rational expectations,} \emph{The Quarterly Journal of Economics},
  131, 1243--1290.

\bibitem[\protect\citeauthoryear{Statman, Thorley, and Vorkink}{Statman
  et~al.}{2006}]{statman2006investor}
\textsc{Statman, M., S.~Thorley, and K.~Vorkink} (2006): \enquote{Investor
  overconfidence and trading volume,} \emph{The Review of Financial Studies},
  19, 1531--1565.

\bibitem[\protect\citeauthoryear{Yildiz}{Yildiz}{2003}]{yildiz2003bargaining}
\textsc{Yildiz, M.} (2003): \enquote{Bargaining without a common prior—an
  immediate agreement theorem,} \emph{Econometrica}, 71, 793--811.

\end{thebibliography}

\newpage
\appendix

\section{Main Appendix}

\subsection{Second-price auction}
\label{Subsection:auction}

Consider a second-price auction, where, like in \cite{atakan2014auctions}, the winner of the auction gets to choose an action that affects the value of the asset. Specifically, the action has a value that depends on her ability to predict a given variable, as in the examples given in the introduction. Formally, fixing the environment defined above (DGP, agents, etc.), consider a game with the following timing:
\begin{enumerate}
\item Nature draws $\theta_0 \in \Theta$;
\item All agents see a common dataset $D_n$ drawn according to $Q_{\theta_0}$;
\item Agents submit bid in a sealed-bid second-price auction;
\item The winner observes $x$ randomly drawn according to $P$ and chooses a real-valued action $a$;
\item The winner gets a lump sum payoff of $M-(y-a)^2$, where $M$ is a large positive number. 
\end{enumerate}

Every bidder seeks to minimize the expected value $M-(y-a)^2$, leading to the expected loss function discussed above.

Because agents see a common data set, an agent with prior $\pi$ has an expected value of $M- L^{*}(\pi, D_n)$ for winning. In the standard dominant equilibrium, the winning agent is the one with the highest value: since $M$ is common across agents, the winner is thus the agent with the lowest expected loss (according to her own prior) given the observed data. Notice that since all agents observe the same dataset, and thus there is no asymmetric information (only heterogenous priors), winner's-curse-type considerations do not apply.\footnote{Our results possibly shed light on political competition/ board meetings. While we do not develop these formally, intuitively, these would correspond to an analogous all-pay auction. Agents have different models of how to forecast payoff-relevant unknowns from covariates. The action taken (by the government body or company) depends on this forecast. Agents' willingness to lobby for their model depends on how confident they are in their model, and the amount of effort they spend lobbying influences selection.}

\subsection{Proof of Lemma \ref{lemma:1agentposteriorloss}}
\label{subsection:ProofLemma1}

\oneagentposteriorloss*

\begin{proof}
Fix a data set $D_n$. We need to analyze
\begin{align*}
&\mathbb{E}_{\pi} \left[ \mathbb{E}_{P} \Big [ (x^{\prime} \beta - f^*_{(\pi,D_n)}(x))^2 \Big ]  \Big| D_n \right].\\
\intertext{Substituting $f^*$ from \eqref{eqn:bayesopt}, we have that this term }
=& \mathbb{E}_{\pi} \left[  \mathbb{E}_{P} \Big [ ( ( \beta- \mathbb{E}_\pi[\beta | D_n] )^{\prime}x )^2 \Big ]  \Big | D_n \right].\\
\intertext{Recalling that for a scalar $a$, $a = \textrm{tr}(a)$, we have }
=&   \mathbb{E}_{\pi} \left[  \mathbb{E}_{P} \Big [ \textrm{tr}[ ( ( \beta- \mathbb{E}_\pi[\beta | D_n] )^{\prime}x )^2] \Big ]  \Big | D_n \right]. \\
\intertext{Then by symmetry and linearity of the trace operator, we can conclude, } 
=&  \mathbb{E}_{\pi} \left[  \mathbb{E}_{P} \Big [ \textrm{tr}[ ( \beta- \mathbb{E}_\pi[\beta | D_n] ) ( \beta- \mathbb{E}_\pi[\beta | D_n] )^{\prime}xx^{\prime} ] \Big ]  \Big | D_n \right],   \\
=&  \mathbb{E}_{\pi} \left[   \textrm{tr}[ ( \beta- \mathbb{E}_\pi[\beta | D_n] ) ( \beta- \mathbb{E}_\pi[\beta | D_n] )^{\prime}\mathbb{E}_{P}[xx^{\prime}] ]   \Big | D_n \right],\\
=&   \textrm{tr} \Big [ \mathbb{E}_{\pi} \left[  ( \beta- \mathbb{E}_\pi[\beta | D_n] ) ( \beta- \mathbb{E}_\pi[\beta | D_n] )^{\prime} \mathbb{E}_{P}[xx^{\prime}] \Big | D_n \right]   \Big ],  \\
\intertext{Since $\beta$ and $P$ are independent under the posterior, we have that,}
& \mathbb{E}_{\pi} \left[  ( \beta- \mathbb{E}_\pi[\beta | D_n] ) ( \beta- \mathbb{E}_\pi[\beta | D_n] )^{\prime} \mathbb{E}_{P}[xx^{\prime}] \Big | D_n \right]
\intertext{equals}
&\mathbb{E}_{\pi} \left[  ( \beta- \mathbb{E}_\pi[\beta | D_n] ) ( \beta- \mathbb{E}_\pi[\beta | D_n] )^{\prime} \Big | D_n \right] \mathbb{E}_{\pi}\left[  \mathbb{E}_{P}[xx^{\prime}] \Big | D_n \right]. 
\intertext{Finally, by the definition of  variance, we have the desired form}
=& \textrm{tr}(  \mathbb{V}_\pi(\beta | D_n) \mathbb{E}_{\pi}\left [ E_{P}[xx^{\prime}] \: | \: D_n \right] ).  \qedhere
\end{align*}
\end{proof}

\subsection{Proof of Lemma \ref{prop:unknownP}}\label{app:proofoflemma2}

Before we delve into the proof, we recall a   few facts about the Normal-Inverse-Gamma-Inverse-Wishart distribution which may be useful. In particular, straightforward algebra shows that given a dataset $D_n$:
 \begin{align}
 \mathbb{E}_\pi[\seps|D_n] &= \frac{\frac{2b_0}{n} + \frac{1}{n} (Y'Y - Y'X_J(X_J'X_J + \gamma_0 |J| \mathbb{I}_{J} )^{-1}X_J'Y)  }{\frac{2 a_0}{n} + 1 - \frac{2}{n} },
 \label{equation:PosteriorMeanSigma}\\
 \mathbb{V}_{\pi}[\beta_{J}|D_n] &= \mathbb{E}_{\pi} \left[\seps | D_n\right] (X_{J}' X_{J} + (\gamma_0 |J|) \mathbb{I}_{|J|})^{-1}. \label{equation:PosterVariance}
 \end{align}
The posterior distribution of $\Sigma_{J(\pi)}$ after observing the data $D_n$ is given as:
    \[ \Sigma_{J(\pi)} | D_n \sim \textrm{Inverse-Wishart}(X'_{J(\pi)}X_{J(\pi)} + \gamma |J(\pi)| \mathbb{I}_{J(\pi)}, n + 2|J(\pi)|+1). \]
    Therefore, 
    \begin{align}\mathbb{E}_{\pi}[\Sigma_{J(\pi)} | D_n]= \frac{X'_{J(\pi)}X_{J(\pi)} + \gamma |J(\pi)| \mathbb{I}_{J(\pi)}}{n + |J(\pi)|}.\label{eqn:PosteriorWishart}
    \end{align}

\unknownP*

\begin{proof}
Note that Remark \ref{ref:rem_independence} implies that the condition of Lemma \ref{lemma:1agentposteriorloss} is satisfied and therefore we the decomposition, i.e. \eqref{equation:PosteriorLoss}:
    \[L^*(\pi, D_n)= \mathbb{E}_{\pi}[\seps | D_n] + \textrm{tr} \left( \mathbb{V}_{\pi}(\beta | D_n) \mathbb{E}_{\pi} [ \Sigma | D_n ]  \right). \]
Note that for any model $J(\pi)$, the mean of $\Sigma_{J(\pi)}$ under the prior is $ \gamma \mathbb{I}_{|J(\pi)|},$ and therefore the prior loss is
    \[ \mathbb{E}_{\pi}[\sigma_{\epsilon}^2]\left( 1 + \textrm{tr}(\mathbb{I}_{|J(\pi)|})/|J(\pi)|  \right) = 2 \mathbb{E}_{\pi}[\seps].  \]
    This means all models have the same ex-ante loss prior to observing the data. 
    Substituting in $V_{\pi}(\beta|D_n)$ from \eqref{equation:PosterVariance}, and $\mathbb{E}_{\pi}[\Sigma_{J(\pi)} | D_n]$ from \eqref{eqn:PosteriorWishart} into \eqref{equation:PosteriorLoss} we have that
    \[L^*(\pi,D_n)= \mathbb{E}_{\pi}[\sigma_{\epsilon}^2 | D_n]\left( 1 + \frac{|J(\pi)|}{n+|J(\pi)|}  \right),  \]
    as desired.
\end{proof}  

\subsection{Proof of Proposition \ref{prop:1dataunknownP}}

\onedataunknownP*

\begin{proof}
Denote the single datapoint as $D_1 = (Y,X)$, where $Y \in \mathbb{R}$ and $X \in \mathbb{R}^{1 \times k}$ ($k$ is the number of covariates), where $X=(x_1,\ldots, x_k)$. First, observe that for any agent $j$ with a single explanatory variable $\kappa$ in his model (denoted $x_{\kappa}$),
\begin{align*}
L^*(\pi_j, D_1) &= \frac{b_0 + \frac12 \left(y^2 - \frac{y^2x_{\kappa}^2}{x_{\kappa}^2 + \gamma}\right) }{a_0 - \frac12} \left(1 + \frac12 \right), \\
&= \frac{b_0 + \frac12  \frac{y^2\gamma}{x_{\kappa}^2 + \gamma} }{a_0 - \frac12} \left(1 + \frac12 \right).
\end{align*}
The winning agent among the single variable models will therefore clearly be the agent with the variable $\kappa$ that maximizes $x^2_{\kappa}$. Without loss of generality, call this variable $1$.

To economize on notation, now consider the full model with all the explanatory variables, it will be clear from the logic that this argument will work for any model larger than a single variable. For an agent $j$ with all $k$ variables, we know that 
\begin{align*}
L^*(\pi_j, D_1) = \frac{b_0 + \frac{y^2}{2} \left(1- X(X'X + \gamma k \mathbb{I}_k )^{-1} X'\right)}{a_0 - \frac12} \left(1 + \frac{k}{n+k} \right).
\end{align*}
We show that this model always loses to the ``best'' single variable model. To do this, it is sufficient to show: 
\begin{align*}
&(1- X(X'X + \gamma k \mathbb{I}_k)^{-1}X') \geq \frac{\gamma}{x_1^2 + \gamma}.
\end{align*}

Algebra shows that
\begin{align*}
&(1- X(X'X + \gamma k \mathbb{I}_k)^{-1}X') \geq \frac{\gamma}{x_1^2 + \gamma},\\
\iff& X(X'X + \gamma k \mathbb{I}_k)^{-1}X' \leq \frac{x_1^2}{x_1^2 + \gamma}.
\end{align*}
Observe that $X(X'X + \gamma k \mathbb{I}_k)^{-1}X'$ is a scalar. We know that for a scalar, $a = \text{tr}(a)$. Therefore:
\begin{align*}
&X(X'X + \gamma k \mathbb{I}_k)^{-1}X',\\
=& \mathrm{tr}[X(X'X + \gamma k \mathbb{I}_k)^{-1}X'],\\
=& \mathrm{tr}[(X'X + \gamma k \mathbb{I}_k)^{-1}X'X],\\
=& \mathrm{tr}\left[(\frac{1}{\gamma k} X'X +  \mathbb{I}_k)^{-1}\frac{1}{\gamma k}X'X \right ].
\intertext{Denote $\frac{1}{\gamma k} X'X $ as $A$. Substituting}
=& \mathrm{tr}[(A +  \mathbb{I}_k)^{-1}A].
\end{align*}
Observe that if $\lambda$ is an eigenvalue of $A$, then $\frac{\lambda}{1+ \lambda}$ is an eigenvalue of $(A +  \mathbb{I}_k)^{-1}A$. To see this, suppose $v$ is an eigenvector of $A$ with eigenvalue $\lambda$. Then, 
\begin{align*}
&A v = \lambda v, \\
\implies& (A+ \mathbb{I}_k) v = (\lambda+1) v ,\\
\implies & (A+ \mathbb{I}_k)^{-1} v = \frac{1}{1+\lambda}v ,\\
\implies & (A+ \mathbb{I}_k)^{-1} A v = \frac{\lambda}{1+\lambda}v .
\end{align*}
Substituting this in, we have
\begin{align*}
\mathrm{tr}[(A +  \mathbb{I}_k)^{-1}A]=& \sum_{i=1}^k \frac{\lambda_i}{1+\lambda_i}.
\end{align*}
Therefore we are left to show that
\begin{align*}
&\sum_{i=1}^k \frac{\lambda_i}{1+\lambda_i} \leq \frac{x_1^2}{x_1^2 + \gamma}
\end{align*}
Here $\lambda_i$'s are the eigenvalues of $\frac{1}{\gamma k} X'X$. This  implies that $\sum_i \lambda_i = \frac{1}{\gamma k} \sum_i x_i^2.$

Note that $X'X$ is not full rank, indeed, its null space is of dimension $k-1$. Therefore it has $k-1$ multiplicity eigenvalue of $0$. The unique non-zero eigenvalue must then be $ \frac{1}{\gamma k} \sum_i x_i^2$. 

Substituting in, we have 
\begin{align*}
\sum_{i=1}^k \frac{\lambda_i}{1+\lambda_i}=& \frac{\frac{1}{\gamma k} \sum_i x_i^2}{\frac{1}{\gamma k} \sum_i x_i^2 + 1},\\
=& \frac{\frac{1}{k} \sum_i x_i^2}{\frac{1}{ k} \sum_i x_i^2 + \gamma},\\
\leq & \frac{x_1^2}{x_1^2 + \gamma}.
\end{align*}
where the last inequality follows since we assumed that $x_1^2 = \max_i \{x_i^2: 1 \leq i \leq k \}$.
\end{proof}

\subsection{Proof of Proposition \ref{prop:highprobwinnerUP} }

\highprobwinnerUP*

\begin{proof}
Fix any prior $\pi \in \Pi$ such that $|J(\pi)| = |\underline{J}|$. Let $\Pi' \subset \Pi$ be the set of priors with size larger than $|\underline{J}|$, i.e. $\Pi' = \{\pi': \pi' \in \Pi \text{ and } |J(\pi')| > |\underline{J}|\}$. 

From \eqref{eqn:compareUP}, we have that for any other prior $\pi' \in \Pi'$:
\begin{align*}
    &L^*(\pi, D_n) \geq L^*(\pi', D_n),\\
    \iff &   \frac{\mathbb{E}_\pi[\seps|D_n]}{ \mathbb{E}_{\pi'}[\seps|D_n] } \geq \frac{\left(1 + \frac{|J(\pi')|}{n + |J(\pi')|}\right)}{\left(1 + \frac{|J(\pi)|}{n + |J(\pi)|}\right)},\\
\intertext{We know from \eqref{equation:PosteriorMeanSigma} that the left hand side,}
&\frac{\mathbb{E}_\pi[\seps|D_n]}{ \mathbb{E}_{\pi'}[\seps|D_n] } = \frac{2b_0 +  (Y'Y - Y'X_{J(\pi)}(X_{J(\pi)}'X_{J(\pi)} + \gamma_0 |J| \mathbb{I}_{J} )^{-1}X_{J(\pi)}'Y)  }{2b_0 +  (Y'Y - Y'X_{J(\pi')}(X_{J(\pi')}'X_{J(\pi')} + \gamma_0 |J| \mathbb{I}_{J} )^{-1}X_{J(\pi')}'Y)}
\end{align*}
Therefore as $b_0$ grows large, we have that (the left hand side) $\frac{\mathbb{E}_\pi[\seps|D_n]}{ \mathbb{E}_{\pi'}[\seps|D_n] }  \to_P 1$. However, the right hand side is a constant that is larger than $1$ by observation. Therefore, for any given $p'$, there exists  $b_0$ large enough so that $\mathbb{P}(D_n: L^*(\pi,D_n) \leq L^*(\pi',D_n)) > p'$. 

Pick $p'$ such that $1-p' \geq \frac{1-p}{|\Pi'|}$. Therefore we have that 
\[ \mathbb{P} \left( D_n: L(\pi, D_n) \leq L(\pi',D_n) \; \forall \pi' \in \Pi' \right) >p. \]
Since $\pi$ is s.t. $J(\pi) = |\underline{J}|$ by assumption, we have the desired result.

\end{proof}

\subsection{Proof of Lemma \ref{lemma:expansion}}

Let the data $D_n = (Y,X)$, $Y \in \Re^n$, $X \in \Re^{n \times k}$ consist of $n$ i.i.d. draws of $y$ and $x$. Let $\mathbb{P}$ denote the true joint distribution of $(y,x)$. 

The density for $Y|X$ corresponding to the Gaussian linear regression model postulated by an agent with variables in $J \subseteq \{1, \ldots, k\}$ is:  
\begin{equation} \label{equation:likelihood}
f(Y | X_{J}; \beta_{J}, \sigma^2 ) := \frac{1}{(2\pi)^{n/2}} \frac{1}{\sigma^{n}} \exp \left( -\frac{1}{ 2 \sigma^2} (Y - X_{J}\beta_{J}  )^{\prime}(Y - X_{J}\beta_{J}  ) \right). 
\end{equation}
Let
\[ \widehat{\beta}_{J} := (X_J'X_J)^{-1}X_J'Y, \quad \widehat{\sigma}^2_J := (Y-X_J'\widehat{\beta}_J)'(Y-X_J'\widehat{\beta}_J)/n,\] 
denote the Maximum Likelihood estimators of $(\beta_J,\sigma^2)$ based on \eqref{equation:likelihood}.  

In what follows, let $\pi$ denote a joint distribution over $(\beta_J, \sigma^2) \in \mathbb{R}^{|J|} \times \mathbb{R}_{+}$. Let $\pi( \cdot | D_n)$ denote the posterior density of  $(\beta_J,\sigma^2)$ based on the likelihood \eqref{equation:likelihood} and the prior $\pi$. Let $\mathbb{E}_{\pi}[\sigma^2 | D_n]$ denote the posterior expectation of $\sigma^2$ under the posterior density.

\begin{lemma} \label{lemma:expansion}
Suppose $\pi$ is a four-times continuously differentiable, strictly positive density function on $(\beta_J, \sigma^2)$. Suppose $\mathbb{E}_{\pi}[\sigma^2 | D_n] < \infty$ almost surely, for $n$ large enough. If $X'X/n$ converges in probability to a positive definite matrix, then
\begin{equation*} \label{equation:KTK_expansion}
       \mathbb{E}_{\pi}[\sigma^2 | D_n] = \widehat{\sigma}_J^2 + \frac{1}{n} \left( 2\widehat{\sigma}^4_J \left \{ \left( \frac{\partial \pi}{\partial \sigma^2} (\widehat{\theta}_J) \right) \cdot \frac{1}{\pi(\widehat{\theta}_J)} \right\} +  {\widehat{\sigma}^2_J} (|J|+4) \right) + O_{\mathbb{P}} \left( \frac{1}{n^2} \right), 
\end{equation*}
where $\widehat{\theta}_J := (\widehat{\beta}'_J,\widehat{\sigma}^2_J)'$.
\end{lemma}

\begin{proof}
The proof has two main steps. First, we introduce some additional notation. Second, we invoke the results of \cite{KTK:1990} and apply them to approximate $\mathbb{E}_{\pi}[\sigma^2 | D_n]$.

\noindent {\scshape Step 0 (Additional Notation):} Let $\theta = (\beta'_J,\sigma^2)'$ and 
\[ h_n(\theta) := -\frac{1}{n} \ln f(Y|X_J); \theta). \] 
The $(i,j)$ component of the matrix of second derivatives of $h_{n}(\theta )$ with respect to $\theta$ (the Hessian of the scaled negative log-likelihood) will be denoted as $h_{ij}(\cdot)$. We omit the dependence on $n$, unless confusion arises. The components of the inverse of the Hessian will be written as $h^{ij}(\cdot)$. Finally, $h_{rsj}(\cdot)$ denotes the partial derivative of $h_{rs}$ with respect to the $j$-th component of $\theta$. \\

\noindent {\scshape Step 1 (Asymptotic Expansion of $\mathbb{E}_{\pi}[\sigma^2 | D_n]$):} \cite{KTK:1990} provide asymptotic expansions for posterior moments of a real-valued function of $\theta$ in terms of the maximizers of the likelihood used to compute the posterior. 

Consider the function
\[ g(\theta) = g((\beta_{J}',\sigma^2)') = \sigma^2. \]

Theorem 4 in \cite{KTK:1990} implies that under the assumptions of our lemma:
\begin{eqnarray*}
\mathbb{E}_{\pi}[g(\theta) | D_n] &=& g(\widehat{\theta}_J) + 
\frac{1}{n} \sum_{1 \leq i,j \leq \textrm{dim}(\theta)} \left( \frac{\partial g }{\partial \theta_i}  (\widehat{\theta}_J) \right) \:  h^{ij}(\widehat{\theta}_J) 
\left \{ \left( \frac{\partial \pi}{\partial \theta_j} (\widehat{\theta}_J ) \right) \cdot \right. \\
&& \left. \frac{1}{\pi(\widehat{\theta}_J)} - \frac{1}{2} \sum_{1 \leq r,s \leq \textrm{dim}(\theta)} h^{rs} (\widehat{\theta}_J) h_{rsj}(\widehat{\theta}_J) \right\} \\
&+& \frac{1}{2n} \sum_{1 \leq i,j \leq \textrm{dim}(\theta)} h^{ij}(\widehat{\theta}_J) \left( \frac{\partial g}{\partial \theta_i \theta_j} (\widehat{\theta}_J) \right) \\ 
&+& O_{\mathbb{P}} \left( \frac{1}{n^2} \right). 
\end{eqnarray*}

\noindent See equation 2.6 in p. 481 of \cite{KTK:1990}.

Since
\[ \frac{\partial g }{\partial \sigma^2}  (\theta)=1, \]
and also
\[\frac{\partial g }{\partial \theta_i}  (\theta) = 0,\] 
for any $i < |J|+1$, the expansion above simplifies to
\begin{eqnarray*}
\mathbb{E}_{\pi}[\sigma^2 | D_n] &=& \widehat{\sigma}^2_J + 
\frac{1}{n} \sum_{1 \leq j \leq |J|+1}  \:  h^{(|J|+1) j}(\widehat{\theta}_J) 
\left \{ \left( \frac{\partial \pi}{\partial \theta_j} (\widehat{\theta} ) \right) \cdot \right. \\
&& \left. \frac{1}{\pi(\widehat{\theta}_J)} - \frac{1}{2} \sum_{1 \leq r,s \leq |J|+1} h^{rs} (\widehat{\theta}_J) h_{rsj}(\widehat{\theta}_J) \right\} \\
&&+ O_{\mathbb{P}} \left( \frac{1}{n^2} \right). 
\end{eqnarray*}
\noindent We now derive explicit formulae for the Hessian matrix, and its inverse elements. The Hessian matrix of $h_n(\theta)$ equals
\begin{equation*}\label{equation:hessian}
\begin{pmatrix}
\frac{1}{n\sigma^2}  {X_{J}}^{\prime} {X_{J}} & \frac{1}{n \sigma^4} {X_{J}}^{\prime}(Y - {X_{J}}'\beta_{J}) \\
\frac{1}{n \sigma^4} (Y - {X_{J}}'\beta_{J})'{X_{J}} & -\frac{1}{2 \sigma^4} + \frac{1}{n \sigma^6}(Y - X_{J}'\beta_J)' (Y - {X_{J}'\beta_J} ),
\end{pmatrix}
\end{equation*}
and the inverse elements of the Hessian evaluated at $\widehat{\theta}$ are:
\begin{equation} \label{equation:inverseHessian}
\begin{pmatrix}
\widehat{\sigma}^2_J  \left( {X_{J}}^{\prime} {X_{J}}/n \right)^{-1} & \mathbf{0} \\
\mathbf{0} & 2 \widehat{\sigma}_J^4,
\end{pmatrix}.
\end{equation}
\noindent This further simplifies the expansion to
\begin{eqnarray*}
    \mathbb{E}_{\pi}[\sigma^2 | D_n] &=& \widehat{\sigma}^2_J +  \frac{2\widehat{\sigma}^4_J}{n} \left \{ \left( \frac{\partial \pi}{\partial \sigma^2} (\widehat{\theta}_J ) \right) \cdot \frac{1}{\pi(\widehat{\theta}_J)} \right. \\
    &&- \left. \frac{1}{2} \sum_{1 \leq r,s \leq |J|+1} h^{rs} (\widehat{\theta}) h_{rs(|J|+1)}(\widehat{\theta}_J) \right\} + O_{\mathbb{P}} \left( \frac{1}{n^2} \right). 
\end{eqnarray*}
\noindent Algebra shows that the matrix that collects the terms $h_{rs( |J|+1)}(\widehat{\theta}_J)$ (which corresponds to the derivative of the Hessian matrix with respect to $\sigma^2$) equals:
\begin{equation} \label{equation:aux_third_derivatives}
\begin{pmatrix}
-\frac{1}{\widehat{\sigma}^4_J }  \left( {X_{J}}^{\prime} {X_{J}}/n \right) & \mathbf{0} \\
\mathbf{0} & - \frac{2}{\widehat{\sigma}_J^6},
\end{pmatrix}.
\end{equation}
The term 
\[\sum_{1 \leq r,s \leq |J|+1} h^{rs} (\widehat{\theta}_J) h_{rs(|J|+1)}(\widehat{\theta}_J),\]
can be written as the sum of all elements of the Hadamard product between the matrices in \eqref{equation:inverseHessian} and \eqref{equation:aux_third_derivatives}. This sum, in turn equals:
\[ \textrm{tr}\left(
\begin{pmatrix}
\widehat{\sigma}^2_J  \left( {X_{J}}^{\prime} {X_{J}}/n \right)^{-1} & \mathbf{0} \\
\mathbf{0} & 2 \widehat{\sigma}_J^4,
\end{pmatrix}
\begin{pmatrix}
-\frac{1}{\widehat{\sigma}^4_J }  \left( {X_{J}}^{\prime} {X_{J}}/n \right) & \mathbf{0} \\
\mathbf{0} & - \frac{2}{\widehat{\sigma}_J^6},
\end{pmatrix}'
\right)
=
-\frac{|J|}{\widehat{\sigma}^2_J } - \frac{4}{ \widehat{\sigma}^2_J }
\]

\noindent We conclude that the Kass-Tierney-Kadane expansion of  $\mathbb{E}_{\pi}[\sigma^2 | D_n]$ equals
\begin{equation*} 
       \mathbb{E}_{\pi}[\sigma^2 | D_n] = \widehat{\sigma}_J^2 + \frac{1}{n} \left( 2\widehat{\sigma}^4_J \left \{ \left( \frac{\partial \pi}{\partial \sigma^2} (\widehat{\theta}_J) \right) \cdot \frac{1}{\pi(\widehat{\theta}_J)} \right\} +  {\widehat{\sigma}^2_J} (|J|+4) \right) + O_{\mathbb{P}} \left( \frac{1}{n^2} \right). \qedhere
\end{equation*}
\end{proof}

\subsection{Proof of Proposition \ref{thm:large_n_winner-easy}}

\largenwinnereasy*

\begin{proof}
Lemma \ref{lemma:1agentposteriorloss} has shown that the (optimized) posterior mean-squared prediction error for an agent with prior $\pi$ is 
\[ L^*(\pi,D_n) = \mathbb{E}_{\pi}[\sigma^2 | D_n] + \textrm{tr} \left( \mathbb{V}_{\pi}(\beta_{J(\pi)}  |  D_n  ) \mathbb{E}_{\pi} \left[ \mathbb{E}_{P}[ x_{J(\pi)} x_{J(\pi)}' ] \: | \:  D_n \right] \right). \]
Under the assumptions of Proposition \ref{thm:large_n_winner-easy}, it follows that for any $\pi \in \Pi$:

\[  L^*(\pi,D_n) = \mathbb{E}_{\pi}[\sigma^2 | D_n] + O_{\mathbb{P}}\left( \frac{1}{n} \right).  \]

Moreover, Assumptions \ref{ass: high-level-easy} and \ref{ass:prior-normalize-easy} imply that the conditions of Lemma \ref{lemma:expansion} are satisfied. Consequently, for any $\pi$ in the finite collection $\Pi$ , the term $\mathbb{E}_{\pi}[\sigma^2 | D_n]$ admits the following \cite{KTK:1990} expansion:
\begin{equation} \label{eqn:expansion_for_proof} 
       \widehat{\sigma}^2(\pi) + O_{\mathbb{P}} \left( \frac{1}{n} \right), 
\end{equation}
where $\widehat{\sigma}^2(\pi)$ denotes the Maximum Likelihood estimator of $\sigma^2_{\epsilon}$ according to the linear regression model with covariates $J(\pi)$. Therefore, for any $\pi \in \Pi$ we have
\begin{equation} \label{eqn:expansion_loss_proof}
    L^*(\pi, D_n) = \widehat{\sigma}^2(\pi) + O_{\mathbb{P}}\left( \frac{1}{n} \right). 
\end{equation}
We will use this expansion to prove the two statements of Theorem \ref{thm:large_n_winner-easy}. 

% The likelihood of the linear regression model with covariates $X_J(\pi)$ satisfies the stochastic Local Asymptotic Normality at $\theta^* \equiv \theta^*(\pi)=(\beta^*(\pi),{\sigma^*}^2(\pi))$.\footnote{Algebra shows the condition is satisfied with

% \begin{eqnarray*}
% V_{\theta^*} & \equiv & 
% \begin{pmatrix}
% \frac{\mathbb{E}_{\mathbb{P}}[x_{J(\pi)}x_{J(\pi)}']}{\sigma^2(\pi)} & \mathbf{0} \\
% \mathbf{0} & \frac{1}{2 \sigma^2(\pi)},
% \end{pmatrix}, \\
% \Delta_{n,\theta^*} &\equiv& V_{\theta^*}^{-1} 
% \begin{pmatrix}
% \frac{1}{{\sigma^*}^2(\pi)} \frac{1}{\sqrt{n}} \sum_{i=1}^{n} x_{J(\pi),i} (y_i-x_{J(\pi),i}'\beta^*(\pi))  \\
% \frac{1}{{\sigma^*}^4(\pi)} \frac{1}{\sqrt{n}} \sum_{i=1}^{n}  (y_i-x_{J(\pi),i}'\beta^*(\pi))^2 - {\sigma^*}^2(\pi)
% \end{pmatrix}.
% \end{eqnarray*}}
% Therefore, Part (3) of Assumption \ref{ass: high-level} implies that Theorem 2.1 in \cite{kleijn2012bernstein} holds and consequently the model estimation uncertainty term equals

% \[  \frac{\sigma^2(\pi)}{n} \textrm{tr}\left( \mathbb{E}_{\mathbb{P}}[ x_{J(\pi)} x_{J(\pi)}' ]^{-1} \mathbb{E}_{P}[ x_{J(\pi)} x_{J(\pi)}' ] \right) + o_{\mathbb{P}}\left( \frac{1}{n} \right).\]

\emph{Misspecified models never win:} We have assumed that the collection $\Pi$ contains a prior $\pi^*$ such that $J_0 \subseteq J(\pi^*)$. This prior defeats any other prior $\pi$ for which $J_0 \nsubseteq J(\pi)$.  To see this, note that Equation \eqref{eqn:expansion_loss_proof} implies
\[ L^*(\pi,D_n)-L^*(\pi^*,D_n)= \widehat{\sigma}^2(\pi) - \widehat{\sigma}^2(\pi^*) + O_{\mathbb{P}}\left( \frac{1}{n} \right). \]

Under Assumption 1, $\widehat{\sigma}^2(\pi^*) \overset{p}{\rightarrow} \sigma^2_0$ (since the larger model includes the relevant covariates). However, since the covariates associated to prior $\pi$ exclude variables that are relevant for prediction 
\[  \widehat{\sigma}^2(\pi) - \widehat{\sigma}^2(\pi^*),  \]
converges in probability to a strictly positive number (the misspecified model has strictly larger residual variance than the true model). This shows that  

$$\lim_{n \rightarrow \infty} \mathbb{P} \left(  \exists \pi \in \argmin_{\pi \in \Pi} L^*(\pi, D_n)  \textrm{ s.t }  J_0 \nsubseteq J(\pi)      \right) = 0.$$

\emph{High-dimensional models win with positive probability:} For the last part of the theorem, let $\pi_L$ denote any prior $\pi$ for which $J_0 \subset J(\pi_L)$ and let $\pi_0$ denote any prior for which $J_0 = J(\pi_0)$. From equation \eqref{eqn:expansion_for_proof}
\[L^*(\pi_0, D_n) - L^*(\pi_L, D_n) = \widehat{\sigma}^2(\pi_0) - \widehat{\sigma}^2(\pi_L) + O_{\mathbb{P}}\left( \frac{1}{n} \right). \] 
Therefore,
\begin{eqnarray*}
\mathbb{P}( L^*(\pi_L, D_n) < L^*(\pi_0, D_n) ) &=&  \mathbb{P}( n( L^*(\pi_0, D_n) - L^*(\pi_L, D_n))>0  ) \\
&=& \mathbb{P} \left( n(\widehat{\sigma}^2(\pi_0) - \widehat{\sigma}^2(\pi_L)  ) > O_{\mathbb{P}}(1)  \right).
\end{eqnarray*}
where we have used the \cite{KTK:1990} expansion in \eqref{eqn:expansion_for_proof}.
Standard algebra of linear regression---e.g., Equation 5.28 in \cite{greene2018econometric} and Theorem 5.1 therein---shows that 
\begin{eqnarray*}
n ( \widehat{\sigma}^2(\pi_0) - \widehat{\sigma}^2(\pi_{L}) )/\sigma^2_0 &\overset{d}{\rightarrow}& \zeta,
\end{eqnarray*}
where $\zeta$ is a chi-squared random variable with $|J(\pi_L)|-|J_0|$ degrees of freedom.

% \footnote{In particular,
% \[\zeta \equiv \xi'[R(\mathbb{E}_{\mathbb{P}}[x_{J(\pi_L)}x_{J(\pi_L)}'])^{-1}R']^{-1} \xi/\sigma^2_0,\] 
% where 
% \[\xi \sim \mathcal{N}_{|J(\pi_L)-J(\pi_0)|}(0,R\mathbb{E}_{\mathbb{P}}[(y-x'\beta_0)^2(x_{J(\pi_{L})}x_{J(\pi_{L})}')^{-1}]R').\]
% In this notation, $R$ is the $|J(\pi_L)-J(\pi_0)| \times |J(\pi_L)|$ matrix that selects the entries of $\beta_{J(\pi_L)}$ that are zero under the model specified by $\pi_0$ and $|J|$ denotes the cardinality of the set $J$.} 

This shows that 
\begin{equation*}
\lim_{n \rightarrow \infty} \mathbb{P}( L^*(\pi_L,D_n) < L^*(\pi_0,D_n) ) \in (0,1]. \qedhere
\end{equation*} 
\end{proof}

\newpage

\section{Supplementary Material}

\subsection{Proof of Proposition \ref{cor:chi-squared-formula}}
\label{subsection:chi-squared-formula}

\begin{restatable}{proposition}{corlargenwinnereasy}\label{cor:chi-squared-formula}
Suppose the conditions of Proposition  \ref{thm:large_n_winner-easy} hold. Suppose, in addition, that for any $\pi$ for which $J_0 \subseteq J(\pi)$,
\[ \textrm{tr} \left( n \mathbb{V}_{\pi}(\beta_{J(\pi)}  |  D_n  ) \mathbb{E}_{\pi} \left[ \mathbb{E}_{P}[ x_{J(\pi)} x_{J(\pi)}' ] \: | \:  D_n \right] \right) \overset{p}{\rightarrow} \sigma^2_0 |J(\pi)|.     \]
Then, for any $\pi, \pi_0$ such that $J(\pi_0) = J_0 \subset J(\pi)$,
\[ \lim_{n \rightarrow \infty} \mathbb{P}\big( L^*(\pi,D_n) < L^*(\pi_0,D_n) \big) \] 
converges to the probability that a chi-squared random variable with $|J(\pi)|-|J_0|$ degrees of freedom exceeds
\begin{equation} \label{eqn:limit}
2(|J(\pi_0)| - |J_0|) + \left( 2\sigma^2_0 \left \{ \left( \frac{\partial \pi}{\partial \sigma^2} (\beta_0,\sigma^2_0) \right) \cdot \frac{1}{\pi(\beta_0,\sigma^2_0)}  - \left( \frac{\partial \pi_0}{\partial \sigma^2} (\beta_0,\sigma^2_0) \right) \cdot \frac{1}{\pi_0(\beta_0,\sigma^2_0)}\right\} \right).
\end{equation}
Moreover, if the marginal distribution over $\sigma^2$ is the same under both $\pi$ and $\pi_0$, the expression in \eqref{eqn:limit} simplifies to
\begin{equation*} 
2(|J(\pi_0)| - |J_0|) + \left( 2\sigma^2_0 \left \{ \left( \frac{\partial \pi_{\beta|\sigma^2}}{\partial \sigma^2} (\beta_0,\sigma^2_0) \right) \cdot \frac{1}{\pi_{\beta|\sigma^2}(\beta_0,\sigma^2_0)}  - \left( \frac{\partial \pi_{0,\beta|\sigma^2}}{\partial \sigma^2} (\beta_0,\sigma^2_0) \right) \cdot \frac{1}{\pi_{0,\beta|\sigma^2}(\beta_0,\sigma^2_0)}\right\} \right).
\end{equation*}

\end{restatable}

\begin{proof}
Under the assumptions of Proposition  \ref{thm:large_n_winner-easy} and the corollary, for any $\pi$ s.t $J_0 \subseteq \pi$:
\[ L^*(\pi,D_n) = \mathbb{E}_{\pi}[\sigma^2 | D_n] + \sigma^2_0 (|J(\pi)|)/n + O_{\mathbb{P}}(1). \]
Lemma \ref{lemma:expansion} and Assumption \ref{ass: high-level-easy} then implies that 
\[L^*(\pi_0,D_n)-L^*(\pi,D_n)\] 
equals
\begin{eqnarray*}
 &=& \widehat{\sigma}^2(\pi_0)-\widehat{\sigma}^2(\pi) \\
&+& \frac{1}{n}\left( 2\sigma^4_0 \left \{ \left( \frac{\partial \pi}{\partial \sigma^2} (\beta_0,\sigma^2_0) \right) \cdot \frac{1}{\pi(\beta_0,\sigma^2_0)}  - \left( \frac{\partial \pi_0}{\partial \sigma^2} (\beta_0,\sigma^2_0) \right) \cdot \frac{1}{\pi_0(\beta_0,\sigma^2_0)}\right\} \right) \\
&+& 2 \sigma^2_0 (|J(\pi_0)|-|J|)/n + O_{\mathbb{P}}\left( \frac{1}{n^2} \right).
\end{eqnarray*}
As argued in Theorem \ref{thm:large_n_winner-easy},
\begin{eqnarray*}
n ( \widehat{\sigma}^2(\pi_0) - \widehat{\sigma}^2(\pi_{L}) )/\sigma^2_0 &\overset{d}{\rightarrow}& \zeta,
\end{eqnarray*}
where $\zeta$ is a chi-squared random variable with $|J(\pi_L)|-|J_0|$ degrees of freedom. The result in \eqref{eqn:limit} then follows. 
To verify the last equation write:
\[ \pi(\beta,\sigma^2) = \pi_{\beta| \sigma^2}(\beta, \sigma^2) \cdot \pi_{\sigma^2}(\sigma^2). \]
Using the chain rule and the fact that the marginal distribution of $\sigma^2$ is the same under both $\pi$ and $\pi_0$ gives the desired result.
\end{proof}

\section{Competing Factor Models: Additional details}

\subsection{Prior hyper-parameters} \label{section:Priors-hyper}
We have assumed that each agent has a prior of the form
\begin{equation}
    \beta | \sigma_{\epsilon}^2 \sim \mathcal{N}(0, (\sigma_{\epsilon}^2/\gamma k) \mathbb{I}_k ), \quad  \sigma_{\epsilon}^2 \sim \textrm{Inv-Gamma}(a_0, b_0).
\end{equation}
In this section we explain how to choose the prior hyper-parameters  $(a_0,b_0,\gamma)$. The parameters $(a_0,b_0)$ are common for all agents, but we allow the parameter $\gamma$ to depend on the agent's relevant covariates. 

\subsubsection{Choosing $\gamma$}
The prior loss for an agent with prior $\pi$ and model $J$ is
\[ \underbrace{\mathbb{E}_{\pi} [ \sigma^2_{\epsilon} ]}_{\textrm{Prior Model Fit}}  + \underbrace{\mathbb{E}_{\pi} [ \sigma^2_{\epsilon} ] \textrm{tr} \left( \mathbb{E}_{P}[x_J x_J^{\prime}] \right) / (\gamma k )}_{\textrm{Prior Model Uncertainty}}.   \]
We take $P$ to equal the empirical distribution of the covariates and set:
\begin{equation} \label{eqn:gamma}
    \gamma = \textrm{tr} \left( \mathbb{E}_{P}[x_Jx_J^{\prime}] \right) / k. 
\end{equation}
This choice of $\gamma$ has two justifications. First, it guarantees that all agents have the same prior loss---provided $(a_0,b_0)$ are common among them. Second, it implies that both model fit and model uncertainty contribute equally to the prior loss.  

\subsubsection{Choosing $(a_0, b_0)$} 
Maximizing the marginal likelihood of the data is a common strategy for choosing hyper-parameters in Bayesian Linear Regression; see for example Chapter 3.5 of \cite{bishop2006pattern}. Let $X$ denote the $n \times k$ matrix containing the $k$ covariates for the $n$ observations in the sample. We fix this matrix and analyze the distribution of 
\begin{equation}
    Y | X, a_0, b_0, \gamma,
\end{equation}
Algebra shows that
\begin{equation}
    Y | X, a_0, b_0, \gamma, \sigma^2 \sim \mathcal{N}_n \left(\mathbf{0}, \sigma^2 \left( \mathbb{I}_n + \frac{X X'}{\gamma k}  \right)  \right).
\end{equation}
Since $\sigma^2 | X, a_0, b_0, \gamma$ is \textrm{Inv-Gamma}$(a_0,b_0)$, one can easily compute the joint distribution of $(Y,\sigma^2) | (X, a_0,b_0, \gamma)$. Marginalizing $\sigma^2$ in such distribution shows that the p.d.f of $Y | X, a_0, b_0, \gamma$ equals:
\begin{equation} \label{eqn:marginal}
 \frac{\textrm{det} \left( \mathbb{I}_n + X X' / \gamma k  \right)^{1/2}}{(2 \pi)^{n/2}} \: \frac{b_0^{a_0}}{\Gamma(a_0)} \frac{\Gamma(a_0 + n/2)}{ b_n ^{a_0 + n/2}  }, 
\end{equation}
where $\Gamma(\cdot)$ is the Gamma function and 
\[ b_n = b_0 + \frac{1}{2} Y' \left( \mathbb{I}_n + \frac{XX'}{\gamma k}  \right)^{-1} Y.   \]
Optimizing (\ref{eqn:marginal}) with respect to $(a_0,b_0)$ is equivalent to maximizing: 
\begin{equation} \label{eqn:obj}
    a_0 \ln(b_0) + \ln \left( \Gamma (a_0 + n/2) \right) -\ln( \Gamma(a_0)) -  \left( a_0 + n/2 \right) \ln \left( b_n \right).
\end{equation}
The first order necessary conditions are
\begin{eqnarray*}
a_0 &:&  \ln(b_0)+ \frac{\Gamma'(a_0+n/2)}{\Gamma(a_0 + n/2)} - \frac{\Gamma'(a_0)}{\Gamma(a_0)} - \ln (b_n) = 0  \\
b_0 &:&  \frac{a_0}{b_0} -\frac{(a_0 + n/2)}{b_n} = 0.
\end{eqnarray*}
A solution to this system of equations must satisfy 
\begin{equation}
a_0 (b_0)  \equiv n \cdot \frac{b_0}{  Y' ( \mathbb{I}_n + XX' / \gamma k )^{-1} Y  }.
\end{equation}
We plug this equation in \eqref{eqn:obj} and optimize numerically with respect to $b_0$.

\subsection{Further Considerations}
\label{subsection:additional_figures}

\subsubsection{Alternative competing models and sample sizes}
\begin{figure}[h!]
\centering
\includegraphics[keepaspectratio, scale=.6]{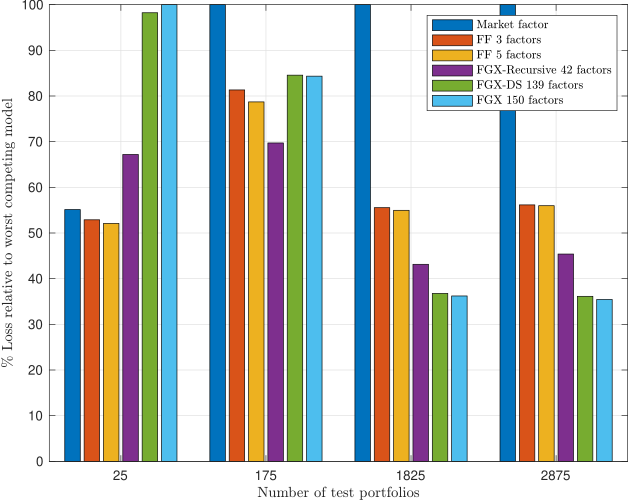}\\
\caption{This figure shows the subjective posterior mean squared prediction error of competing models relative to worst model, by number of test portfolios. FF 3 factors refer to the excess market return, Small Minus Big, and High Minus Low factors \citep{fama1993common}, FF 5 factors refer to the FF 3 factors, Conservative Minus Aggressive, and Robust Minus Weak factors \cite{fama2015five}, FGX-Recursive 42 factors are the factors selected in \cite{feng2020taming} (Section C) using a recursive selection procedure, FGX-DS 139 factors are the benchmark factors selected by \cite{feng2020taming} using the double-selection method, and FGX 150 factors are all the factors in the factor library of \cite{feng2020taming}.}
\label{fig:CFM_all}
\end{figure}

Figure \ref{fig:CFM} presented the results of a competition between three different models with two different sample sizes. Figure \ref{fig:CFM_all} enriches the baseline comparison in two different dimensions. 

The first dimension is to allow for six models. We add the market model \citep{jensen1972capital}; the five-factor model recently suggested by \cite{fama2015five} (which, relative to the three-factor model adds a `robust minus weak' profitability factor and a `conservative minus aggressive' investment factor);  and a 42-factor model selected using the recursive double-selection procedure in \cite{feng2020taming}. 

The second dimension is to allow for two additional sample sizes: 175 ($5 \times 5$) bivariate-sorted portfolios and 1,825 ($5 \times 5$) bivariate-sorted portfolios that sort based on the subset of the 115 factors that have at least 10 stocks on each quintile cell.\footnote{The use of the 175 ($5 \times 5$) bivariate-sorted portfolios is standard in the literature. They are obtained by doing bivariate sorting using size and each of the following seven variables: book-to-market ratio, Market Beta, ``robust minus weak'', ``conservative minus aggressive'', 1-month momentum, 6-month momentum, and 36-month momentum.} 

The results of the competition are consistent with what we report in Figure \ref{fig:CFM}, but with some caveats. The low-dimensional models (3 and 5 factors) still perform better than the high-dimensional models (139 and 150 factors) with small samples ($N=25$ and $N=175$). However, the \cite{fama2015five} five-factor model has a slight edge over the three factor model and market model. This is consistent with the simulations results reported in Figure \ref{fig:CFM} in the paper. Also, we note that with 175 data points, all the competing models have a similar subjective posterior mean-squared prediction error.

\subsubsection{Randomizing selection of portfolios}

In our analysis thus far, we have used the 25 bivariate-sorted portfolios on size and book-to-market of \cite{fama1993common} as our small sample size. While this reflects the standard choice of test assets in the literature, the superior performance of the three and five factor models relative to high-dimensional models may be specific to the set of test assets considered. \cite{kozak2018interpreting} observes that the Fama and French three factors are similar to the first three principal components of the 25 size and book-to-market portfolios. The three and five factor models may perform well on the small sample size simply because they adequately summarize cross-sectional variation in the 25 size and book-to-market portfolios.

\begin{figure}[h!]
\centering
\includegraphics[keepaspectratio, scale=.6]{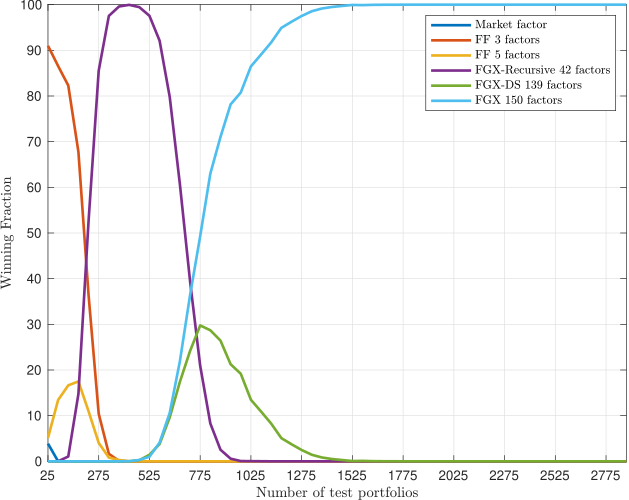}\\
\caption{This figure shows winning rates for the competing models over different sample sizes from $n = 25$ to $2875$. For each sample size, we construct 5,000 simulated datasets by randomly sampling portfolios without replacement up to the given sample size. The winning rate of a model is computed as the fraction of times the model achieves the lowest subjective posterior mean squared prediction error among all competing models.}
\label{fig:winning_rates}
\end{figure}

To  examine the robustness of our findings to the choice of test asset portfolios, we construct 5,000 simulated datasets, each dataset consisting of randomly chosen portfolios up to a given sample size. For each sample size, we then compute the fraction of times a model achieves the lowest subjective posterior mean-squared forecast error. Figure \ref{fig:winning_rates} shows the simulation results. We find that the results remain robust to the choice of test asset portfolios. When $n=25$, the Fama and French three-factor model prevails $90$ percent of the time. As $n$ grows, we see ``waves'' of larger models performing better, with the 42 factor model out-performing other models for sample sizes between $n=225$ and $750$ and with the largest models ($150$ factors) prevailing when $n$ is larger than $1,525$.

\subsubsection{Competing models over time}
\begin{figure}[h!]
\centering
\includegraphics[keepaspectratio, scale=.6]{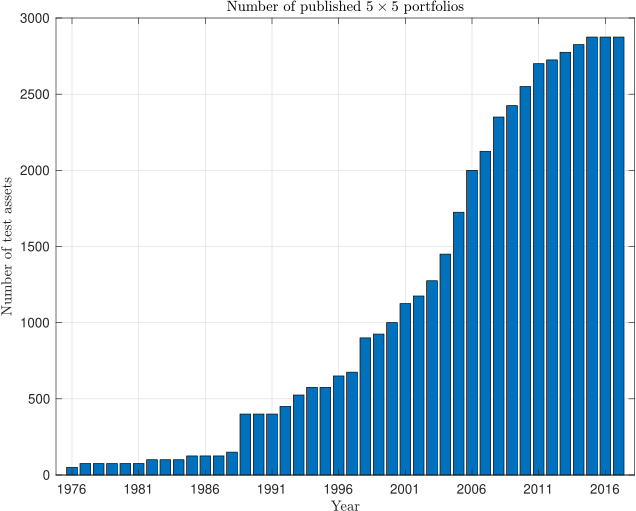}\\
\caption{This figure reports the number of $5 \times 5$ bivariate-sorted portfolios available as test assets each year for the sample period from 1976 to 2017, based on factors from the \cite{feng2020taming} factor library that are published up to the given year.}
\label{fig:nassets_year}
\end{figure}

The test portfolios used in the exercises above are all constructed by sorting on size and some other factor. Since we have the publication data for each factor, we can easily describe the evolution of the number of available test portfolios over time. We report this in Figure \ref{fig:nassets_year}, starting from 1976.   

We also conduct two additional exercises. First, we consider a sample of only the 25 $(5 \times 5)$ bivariate-sorted portfolios, but consider competition between the three-factor model and the factor zoo available each year. Second, we consider a similar competition, but now we assume that the available sample consists of all portfolios published up to a given year. With the small sample, the three-factor model consistently performs better than the factor zoo, with relative performance improving as the size of the factor zoo increases over time. Conversely, when evaluating the models on all available portfolios up to a given time, we see that the factor zoo mostly performs better than the three-factor model. We note that in 1976, both models have similar subjective posterior mean-squared prediction error, albeit with a slight edge for the three-factor model.

\begin{figure}[h!]
\centering
\includegraphics[keepaspectratio, scale=.65]{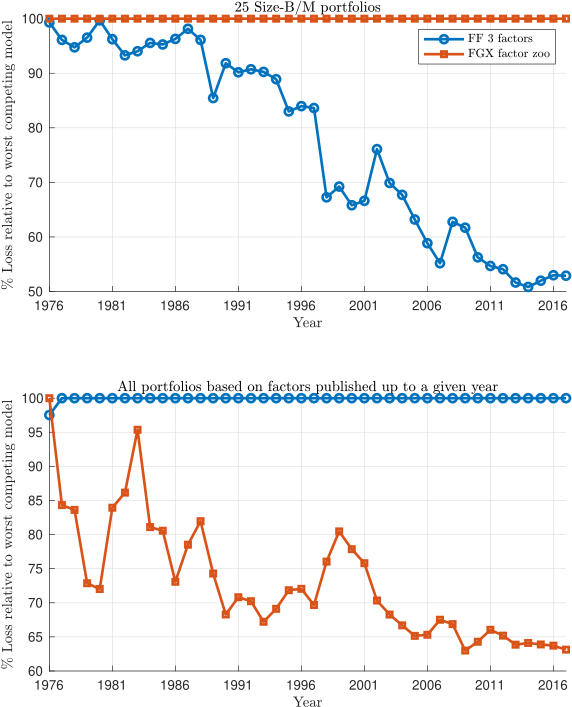}\\
\caption{Subject posterior mean squared prediction error of competing models relative to the worst model, over the sample period from 1976 to 2017. For each year, we compute the relative loss using data on returns available up to the given year. The top panel reports the relative loss of the competing models evaluated on 25 size and book-to-market bivariate sorted portfolios, whereas the bottom panel reports the relative loss evaluated on the size and book-to-market bivariate sorted portfolios and $5 \times 5$ portfolios based on factors in the \cite{feng2020taming} factor library published up given year. The FF3 model includes the market, size, and book-to-market ratio factors. The FGX model includes the FF3 factors as well as all other factors published up to a given year.}
\label{fig:CFM_year}
\end{figure}

\newpage

\subsection{Additional Tables}
\label{app:c3}

\begin{table}[!ht]
\begin{filecontents*}{figures/portfolio_sample.txt}
Factor library of \cite{feng2020taming}	150	
Factors with 5x5 bivariate-sorted portfolios available	115	2875
Factors with 5x5 bivariate-sorted portfolios with at least 10 stocks	73	1825
Selected 5x5 bivariate-sorted portfolios from Kenneth French's website	7	175
5x5 portfolios sorted by size and book-to-market ratio	1	25
\end{filecontents*}
\csvreader[ 
    longtable=p{7.5cm} c c,
    table head= \toprule
                \bfseries Filtering criteria & 
                \bfseries Number.Factors & 
                \bfseries Number.Portfolios \\
                \midrule,
    table foot= \bottomrule,
    no head,
    column count = 3,
    separator=tab,
    before reading=
    \caption*{\noindent This table reports the filtering criteria used to select the sample of portfolios. The second column (Number.Factors) states the number of long-short factors that can be constructed from the selected portfolios, and the third column (Number.Portfolios) states the corresponding number of portfolios chosen after the criteria is applied. Selected portfolios from Kenneth French's website includes 25 potfolios sorted by size and book-to-market ratio, 25 portfolios sorted by size and beta, 25 portfolios sorted by size and operating profitability, 25 portfolios sorted by size and investment, 25 portfolios sorted by size and short-term reversal on prior (1-1) return, 25 portfolios sorted by size and momentum on prior (2-12) return, and 25 portfolios sorted by size and long-term reversal on prior (13-60) return. }
    ]{figures/portfolio_sample.txt}{1=\aa, 2=\bb, 3=\cc}{\aa & \bb & \cc \\}
\end{table}

\newpage

\newcommand\boldID[2]{\ifthenelse{\equal{#2}{NA}}{#1}{#1$^*$}}
\newgeometry{left=1in, right=1in, bottom=1in, top=1in}
\begin{landscape}

\small
\setlength\extrarowheight{-3pt}

\begin{filecontents*}{figures/factor_zoo.csv}
1,Excess Market Return,NA,0.64%,1972,Black and Jensen and Scholes
2,Market Beta,0,-0.08%,1973,Fama and Macbeth
3,Earnings to price,1,0.28%,1977,Basu
4,Dividend to price,NA,0.01%,1979,Litzenberger and Ramaswamy
5,Unexpected quarterly earnings,1,0.12%,1982,Rendelman and Jones and Latane
6,Share price,NA,0.02%,1982,Miller and Scholes
7,Long-Term Reversal,1,0.34%,1985,De Bondt and Thaler
8,Leverage,0,0.21%,1988,Bhandari
9,Cash flow to debt,0,-0.09%,1989,Ou and Penman
10,Current ratio,0,0.06%,1989,Ou and Penman
11,% change in current ratio,1,0.00%,1989,Ou and Penman
12,% change in quick ratio,1,-0.04%,1989,Ou and Penman
13,% change sales-to-inventory,1,0.17%,1989,Ou and Penman
14,Quick ratio,1,-0.02%,1989,Ou and Penman
15,Sales to cash,1,0.01%,1989,Ou and Penman
16,Sales to inventory,1,0.09%,1989,Ou and Penman
17,Sales to receivables,1,0.14%,1989,Ou and Penman
18,Bid-ask spread,0,-0.04%,1989,Amihud and Mendelson
19,Depreciation / PP&E,1,0.11%,1992,Holthausen and Larcker
20,% change in depreciation,1,0.08%,1992,Holthausen and Larcker
21,Small Minus Big,NA,0.21%,1993,Fama and French
22,High Minus Low,1,0.28%,1993,Fama and French
23,1-month momentum,1,0.15%,1993,Jegadeesh and Titman
24,6-month momentum,0,0.21%,1993,Jegadeesh and Titman
25,36-month momentum,0,0.09%,1993,Jegadeesh and Titman
26,Sales growth,1,0.04%,1994,Lakonishok and Shleifer and Vishny
27,Cash flow-to-price,1,0.31%,1994,Lakonishok and Shleifer and Vishny
28,New equity issue,NA,0.10%,1995,Loughran and Ritter
29,Dividend initiation,NA,-0.03%,1995,Michaely and Thaler and Womack
30,Dividend omission,NA,-0.18%,1995,Michaely and Thaler and Womack
31,Working capital accruals,1,0.22%,1996,Sloan
32,Sales to price,0,0.35%,1996,Barbee and Mukherji and Raines
33,Capital turnover,1,-0.11%,1996,Haugen and Baker
34,Momentum,1,0.63%,1997,Carhart
35,Share turnover,0,-0.02%,1998,Datar and Naik and Radcliffe
36,% change in gross margin - % change in sales,1,-0.05%,1998,Abarbanell and Bushee
37,% change in sales - % change in inventory,1,0.14%,1998,Abarbanell and Bushee
38,% change in sales - % change in A/R,1,0.14%,1998,Abarbanell and Bushee
39,% change in sales - % change in SG&A,1,0.09%,1998,Abarbanell and Bushee
40,Effective Tax Rate,0,-0.04%,1998,Abarbanell and Bushee
41,Labor Force Efficiency,1,-0.03%,1998,Abarbanell and Bushee
42,Ohlson's O-score,NA,0.05%,1998,Dichev
43,Altman's Z-score,0,0.20%,1998,Dichev
44,Industry adjusted % change in capital expenditures,1,0.10%,1998,Abarbanell and Bushee
45,Number of earnings increases,NA,0.01%,1999,Barth and Elliott and Finn
46,Industry momentum,1,0.01%,1999,Moskowitz and Grinblatt
47,Financial statements score,NA,0.08%,2000,Piotroski
48,Industry-adjusted book to market,0,0.22%,2000,Asness and Porter and Stevens
49,Industry-adjusted cash flow to price ratio,1,0.26%,2000,Asness and Porter and Stevens
50,Industry-adjusted change in employees,1,-0.01%,2000,Asness and Porter and Stevens
51,Industry-adjusted size,NA,0.36%,2000,Asness and Porter and Stevens
52,Dollar trading volume,NA,0.38%,2001,Chordia and Subrahmanyam and  Anshuman
53,Volatility of liquidity (dollar trading volume),NA,0.20%,2001,Chordia and Subrahmanyam and Anshuman
54,Volatility of liquidity (share turnover),0,0.02%,2001,Chordia and Subrahmanyam andAnshuman
55,Advertising Expense-to-market,0,-0.13%,2001,Chan and Lakonishok and Sougiannis
56,R&D Expense-to-market,0,0.34%,2001,Chan and Lakonishok and Sougiannis
57,R&D-to-sales,1,0.06%,2001,Chan and Lakonishok and Sougiannis
58,Kaplan-Zingales Index,0,0.22%,2001,Lamont and Polk and Saa-Requejo
59,Change in inventory,1,0.18%,2002,Thomas and Zhang
60,Change in tax expense,1,0.09%,2002,Thomas and Zhang
61,Illiquidity,NA,0.34%,2002,Amihud
62,Liquidity,NA,0.38%,2003,Pastor and Stambaugh
63,Idiosyncratic return volatility,0,0.07%,2003,Ali and Hwang and Trombley
64,Growth in long term net operating assets,1,0.22%,2003,Fairfield and Whisenant and Yohn
65,Order backlog,0,0.05%,2003,Rajgopal and Shevlin and Venkatachalam
66,Changes in Long-term Net Operating Assets,1,0.24%,2003,Fairfield and Whisenant and Yohn
67,Cash flow to price ratio,1,0.27%,2004,Desai and Rajgopal and Venkatachalam
68,R&D increase,NA,0.06%,2004,Eberhart and Maxwell and Siddique
69,Corporate investment,1,0.13%,2004,Titman and Wei and Xie
70,Earnings volatility,1,0.10%,2004,Francis and LaFond and Olsson and Schipper
71,Abnormal Corporate Investment,1,0.13%,2004,Titman and Wei and Xie
72,Net Operating Assets,1,0.31%,2004,Hirshleifer and Hou and Teoh and Zhang
73,Changes in Net Operating Assets,1,0.14%,2004,Hirshleifer and Hou and Teoh and Zhang
74,Tax income to book income,1,0.14%,2004,Lev and Nissim
75,Price delay,0,0.07%,2005,Hou and Moskowitz
76,years since first Compustat coverage,NA,0.01%,2005,Jiang and Lee and Zhang
77,Growth in common shareholder equity,1,0.15%,2005,Richardson and Sloan and Soliman and Tuna
78,Growth in long-term debt,1,0.06%,2005,Richardson and Sloan and Soliman and Tuna
79,Change in Current Operating Assets,1,0.19%,2005,Richardson and Sloan and Soliman and Tuna
80,Change in Current Operating Liabilities,1,0.03%,2005,Richardson and Sloan and Soliman and Tuna
81,Changes in Net Non-cash Working Capital,NA,0.11%,2005,Richardson and Sloan and Soliman and Tuna
82,Change in Non-current Operating Assets,0,0.21%,2005,Richardson and Sloan and Soliman and Tuna
83,Change in Non-current Operating Liabilities,1,0.04%,2005,Richardson and Sloan and Soliman and Tuna
84,Change in Net Non-current Operating Assets,0,0.23%,2005,Richardson and Sloan and Soliman and Tuna
85,Change in Net Financial Assets,0,0.23%,2005,Richardson and Sloan and Soliman and Tuna
86,Total accruals,NA,0.19%,2005,Richardson and Sloan and Soliman and Tuna
87,Change in Short- term Investments,NA,-0.03%,2005,Richardson and Sloan and Soliman and Tuna
88,Change in Financial Liabilities,1,0.18%,2005,Richardson and Sloan and Soliman and Tuna
89,Change in Book Equity,0,0.17%,2005,Richardson and Sloan and Soliman and Tuna
90,Financial statements score,NA,0.17%,2005,Mohanram
91,Change in 6-month momentum,0,0.21%,2006,Gettleman and Marks
92,Growth in capital expenditures,1,0.14%,2006,Anderson and Garcia-Feijoo
93,Return volatility,0,-0.02%,2006,Ang and Hodrick and Xing and Zhang
94,Zero trading days,0,-0.05%,2006,Liu
95,Three-year Investment Growth,1,0.11%,2006,Anderson and Garcia-Feijoo
96,Composite Equity Issuance,0,-0.01%,2006,Daniel and Titman
97,Net equity finance,0,0.08%,2006,Bradshaw and Richardson and Sloan
98,Net debt finance,1,0.17%,2006,Bradshaw and Richardson and Sloan
99,Net external finance,1,0.22%,2006,Bradshaw and Richardson and Sloan
100,Revenue Surprises,1,0.05%,2006,Jegadeesh and Livnat
101,Industry Concentration,1,0.03%,2006,Hou and Robinson
102,Whited-Wu Index,NA,-0.02%,2006,Whited and Wu
103,Return on invested capital,0,0.18%,2007,Brown and Rowe
104,Debt capacity/firm tangibility,1,0.05%,2007,Almeida and Campello
105,Payout yield,0,0.16%,2007,Boudoukh and Michaely and Richardson and Roberts
106,Net payout yield,0,0.16%,2007,Boudoukh and Michaely and Richardson and Roberts
107,Net debt-to-price,NA,0.02%,2007,Penman and Richardson and Tuna
108,Enterprise book-to-price,0,0.14%,2007,Penman and Richardson and Tuna
109,Change in shares outstanding,NA,0.24%,2008,Pontiff and Woodgate
110,Abnormal earnings announcement volume,1,-0.08%,2008,Lerman and Livnat and Mendenhall
111,Earnings announcement return,1,0.02%,2008,Kishore and Brandt and Santa-Clara and Venkatachalam
112,Seasonality,0,0.16%,2008,Heston and Sadka
113,Changes in PPE and Inventory-to-assets,1,0.19%,2008,Lyandres and Sun and Zhang
114,Investment Growth,1,0.17%,2008,Xing
115,Composite Debt Issuance,1,0.08%,2008,Lyandres and Sun and Zhang
116,Return on net operating assets,NA,0.09%,2008,Soliman
117,Profit margin,0,0.02%,2008,Soliman
118,Asset turnover,NA,0.06%,2008,Soliman
119,Industry-adjusted change in asset turnover,1,0.14%,2008,Soliman
120,Industry-adjusted change in profit margin,1,-0.01%,2008,Soliman
121,Cash productivity,1,0.27%,2009,Chandrashekar and Rao
122,Sin stocks,NA,0.44%,2009,Hong and Kacperczyk
123,Revenue surprise,0,0.12%,2009,Kama
124,Cash flow volatility,0,0.20%,2009,Huang
125,Absolute accruals,1,-0.05%,2010,Bandyopadhyay and Huang and Wirjanto
126,Capital expenditures and inventory,1,0.19%,2010,Chen and Zhang
127,Return on assets,0,-0.09%,2010,Balakrishnan and Bartov and Faurel
128,Accrual volatility,0,0.19%,2010,Bandyopadhyay and Huang and Wirjanto
129,Industry-adjusted Real Estate Ratio,0,0.11%,2010,Tuzel
130,Percent accruals,1,0.16%,2011,Hafzalla and Lundholm and Van Winkle
131,Maximum daily return,0,0.00%,2011,Bali and Cakici and Whitelaw
132,Operating Leverage,1,0.20%,2011,Novy-Marx
133,Inventory Growth,1,0.13%,2011,Belo and Lin
134,Percent Operating Accruals,1,0.15%,2011,Hafzalla and Lundholm and Van Winkle
135,Enterprise multiple,0,0.11%,2011,Loughran and Wellman
136,Cash holdings,1,0.13%,2012,Palazzo
137,HML Devil,NA,0.23%,2013,Asness and Frazzini
138,Gross profitability,1,0.15%,2013,Novy-Marx
139,Organizational Capital,1,0.21%,2013,Eisfeldt and Papanikolaou
140,Betting Against Beta,NA,0.91%,2014,Frazzini and Pedersen
141,Quality Minus Junk,NA,0.43%,2014,Asness and Frazzini and Pedersen
142,Employee growth rate,1,0.08%,2014,Bazdresch and Belo and Lin
143,Growth in advertising expense,0,0.07%,2014,Lou
144,Book Asset Liquidity,NA,0.09%,2014,Ortiz-Molina and Phillips
145,Robust Minus Weak,1,0.34%,2015,Fama and French
146,Conservative Minus Aggressive,1,0.26%,2015,Fama and French
147,HXZ Investment,NA,0.34%,2015,Hou and Xue and Zhang
148,HXZ Profitability,NA,0.57%,2015,Hou and Xue and Zhang
149,Intermediary Risk Factor,NA,0.15%,2016,He and Kelly and Manela
150,Convertible debt indicator,NA,0.11%,2016,Valta
\end{filecontents*}
\csvreader[
    no head,
    longtable= l p{8cm} c S c p{7cm}, 
    column count=6, 
    table head= \toprule 
                \bfseries ID & 
                \bfseries Description & 
                \bfseries $N\geq10$ & 
                \bfseries Avg.Ret & 
                \bfseries Year.Pub & 
                \bfseries Authors \\ 
                \midrule \endhead 
                \bottomrule\endfoot,
    respect all = true,
    before reading= 
    \captionof*{table}{\noindent This table replicates the list of 150 factors and various descriptive statistics from the factor library in the Appendix of \cite{feng2020taming}. Of the 150 factors, 15 factors come from publicly available sources from the respective authors' websites, and the remaining 135 factors are constructed from long-short value weighted $3 \times 2$ bivariate-sorted portfolios. See \cite{feng2020taming} Section II.A.1 for details on factor construction. Asterisks on IDs indicate the 115 factors where $5 \times 5$ bivariate sorted portfolios are also available. The values in column $N \geq 10$ equal 1 if there are at least 10 stocks in each of the $5 \times 5$ bivariate-sorted portfolios, 0 if there are less than 10 in any of the portfolios, and NA if data is not available. The column Avg.Ret is the monthly average return of the traded factors, Year.Pub is the year of publication, and Authors are the authors of the respective papers.}\label{table:factor_zoo}
    ]{figures/factor_zoo.csv}{1=\aa, 2=\bb, 3=\cc, 4=\dd, 5=\ee, 6=\ff}{\boldID{\aa}{\cc} & \bb & \cc & \dd & \ee & \ff}
\end{landscape}
\restoregeometry

\end{document}